\documentclass[1p]{elsarticle}

\usepackage{lineno,hyperref}
 
\journal{Signal Processing}  

\usepackage{amsmath,graphicx}
 
\usepackage{subfigure}
\usepackage{slashbox}
\usepackage{marginnote}
\usepackage{graphicx,amssymb,amstext,amsmath}
\usepackage{float}
\usepackage{amsfonts}
\usepackage{bbm}
\usepackage{bm}
\usepackage{verbatim}
\usepackage{multirow}
\usepackage{makecell}
\usepackage{color}
\usepackage{caption}
\usepackage{amsthm}
\usepackage{tikz}
\usetikzlibrary{arrows}
\usepackage{pdflscape}

\newcommand{\bmu}{\bm \mu}

\newcommand{\x}{{\bf x}}
\newcommand{\z}{{\bf z}}
\newcommand{\X}{{\bf X}}
\newcommand{\bC}{{\bf C}}
\newcommand{\bI}{{\bf I}}

\newcommand{\cblack}{\textcolor{black}}
\newcommand{\normalized}{\tilde}

\newcommand{\bfootnote}[1]{\cblack{\footnote{\cblack{#1}}}}

\newtheorem{theorem}{Theorem}

\newtheorem{proposition}[theorem]{Proposition}

\begin{document}

\begin{frontmatter}

\title{Improving Population Monte Carlo: Alternative Weighting and Resampling Schemes}

\author{V\'ictor Elvira}
\address{Dep. of Signal Theory and Communic., Universidad Carlos III de Madrid, Legan\'es (Spain).}
\author{Luca Martino}
\address{Dep. of Mathematics and Statistics, University of Helsinki, Helsinki (Finland).}
\author{David Luengo}
\address{ Dep. of Signal Theory and Communic.,Universidad Polit\'ecnica de Madrid, Madrid (Spain).}
\author{M\'onica F. Bugallo}
\address{Dep. of Electrical and Computer Eng., Stony Brook University,  NY (USA).}

\begin{abstract}
Population Monte Carlo (PMC) sampling methods are powerful tools for approximating distributions of static unknowns given a set of observations. These methods are iterative in nature: at each step they generate samples from a proposal distribution and assign them weights according to the importance sampling principle. Critical issues in applying PMC methods are the choice of the generating functions for the samples and the avoidance of the sample degeneracy. In this paper, we propose three new schemes that considerably improve the performance of the original PMC formulation by allowing for better exploration of the space of unknowns and by selecting more adequately the surviving samples. A theoretical analysis is performed, proving the superiority of the novel schemes in terms of variance of the associated estimators and preservation of the sample diversity. Furthermore, we show that they outperform other state of the art algorithms (both in terms of mean square error and robustness w.r.t. initialization) through extensive numerical simulations. 
\end{abstract}


\end{frontmatter}


\section{Introduction}
\label{sec_intro}
{\color{black}
Bayesian signal processing, which has become very popular over the last years in statistical signal processing, requires computing distributions of unknowns conditioned on observations (and moments of them). Unfortunately, these distributions are often impossible to obtain analytically in many real-world challenging problems. 
An alternative is then to resort to Monte Carlo (MC) methods, which approximate the target distributions with random measures composed of samples and associated weights  \cite{Robert04}. 

A well-known class of MC methods are those based on the adaptive importance sampling (AIS) mechanism, such as Population Monte Carlo (PMC) algorithms \cite{Iba01,Cappe04}, which have been used in missing data, tracking, and biological applications, among others \cite{Celeux06,Bink08,Bi09,Barter09,Bhaskar08}.
In these methods, a population of probability density functions (pdfs) is adapted for approximating a target distribution through an iterative importance sampling procedure. 
AIS is often preferred to other MC schemes, such as Markov Chain Monte Carlo (MCMC), since they present several advantages. 
On the one hand, all the generated samples are employed in the estimation {(e.g., there is no ``burn-in'' period)}. 
On the other hand, the corresponding adaptive schemes are more flexible, since they present {fewer} theoretical issues than adaptive MCMC algorithms.
{Namely, the convergence of AIS methods can usually be guaranteed under mild assumptions regarding the tails of the distributions and the stability of the adaptive process, whereas adaptive MCMC schemes must be designed very carefully, since the adaptation procedure can easily jeopardize the ergodicity of the chain (e.g., see \cite{andrieu2008tutorial} or \cite[Section 7.6.3]{Robert04}).}

The most characteristic feature in PMC  \cite{Cappe04} is arguably the use of resampling procedures for adapting the proposal pdfs (see for instance \cite{li2015resampling} for a review of resampling methods in particle filtering). 
The resampling step is a fast, often dimensionality-free, and easy way of adapting the proposal pdfs by using information about the target. 
However, resampling schemes present some important drawbacks, such as the sample impoverishment. At the resampling step, the proposal pdfs with poor performance (i.e., with low associated weights) are likely to be removed, thus yielding a reduction of diversity.
Since the publication of the standard PMC \cite{Cappe04}, several variants have been considered, partly in an attempt to mitigate this issue. In the D-kernel algorithm \cite{Douc07a,Douc07b}, the PMC kernel is a mixture of different kernels and the weights of the mixture are iteratively adapted in an implicit expectation-maximization (EM) algorithm. This procedure is refined through a double Rao-Blackwelization in \cite{iacobucci2010variance}.
The mixture population Monte Carlo algorithm (M-PMC) proposed in \cite{Cappe08} also adapts a mixture of proposal pdfs (weights and parameters of the kernels). The M-PMC belongs to the family of AIS methods, since it iteratively draws the samples from the mixture that is updated at every iteration without an explicit resampling step. Since drawing from the mixture can be interpreted as an implicit multinomial resampling, this method retains some similarities with the standard PMC scheme.
A nonlinear transformation of the importance weights in the PMC framework has also been proposed in \cite{koblents2013population}.
Other sophisticated AIS schemes, such as the AMIS \cite{CORNUET12} and the APIS \cite{APIS15} algorithms, have been recently proposed in the literature.

In this paper, we study three novel PMC schemes that improve the performance of  standard PMC approach by allowing a better exploration of the space of unknowns and by reducing the variance of the estimators. {These alternatives can be applied within some other sophisticated AIS approaches as well, such as the SMC samplers \cite{del2006sequential}}. For this reason, we mainly compare them with the standard PMC  \cite{Cappe04}, since the novel schemes could be automatically combined with the more sophisticated AIS techniques.

First of all, we introduce an alternative form of the importance weights, using a mixture of the proposal pdfs in the denominator of the weight ratio. We provide an exhaustive theoretical analysis, proving the unbiasedness and consistency of the resulting estimator, and showing the reduction in the variance of the estimator w.r.t. the estimator obtained using the standard weights.
We also prove that the use of this mixture decreases the averaged mismatch between the numerator (target) and the function in the denominator of the IS weight in terms of $L_2$ distance. Moreover, we test this alternative scheme in different numerical simulations, including an illustrative toy example in Section \ref{sec_ex1}, showing its practical benefit.

In the second proposed scheme, we generate several samples from every proposal pdf (not only one, as in PMC) and then we resample them jointly (all the samples at once, keeping fixed the total number of proposal pdfs). In the third proposed scheme, we consider again the generation of several samples from every proposal pdf, but the resampling is performed separately on the set of samples coming from each proposal, therefore guaranteeing that there will be exactly one representative from each of the individual {mixture components} in the random measure.  

We show, through extensive computer simulations in several different scenarios, that the three newly proposed variants provide a substantial improvement compared to the standard PMC. {In addition, we test the proposed variants on a standard implementation of the SMC samplers \cite{del2006sequential}, showing also an improvement of the performance.} On the one hand, they yield unbiased estimators with a reduced variance, as also proved theoretically. On the other hand, they outperform the standard PMC in terms of preservation of sample diversity and robustness w.r.t initialization and parameter choice.  
}
\section{Problem Statement}
\label{sec_statement}

Let us consider the variable of interest, $\x\in \mathbb{R}^{D_x}$, and let ${\bf y} \in \mathbb{R}^{D_y}$ be the observed data.
In a Bayesian framework, the posterior probability density function (pdf), here referred as \emph{target}, contains all the information about the parameters of interest and is defined as
\begin{equation}
	\normalized{\pi}(\x| {\bf y})= \frac{\ell({\bf y}|\x) {p_0}(\x)}{Z({\bf y})} \propto \pi(\x) = \ell({\bf y}|\x) {p_0}(\x),
\label{eq:posterior}
\end{equation}
where $\ell({\bf y}|\x)$ is the likelihood function, ${p_0}(\x)$ is the prior pdf, and $Z({\bf y})$ is the model evidence or partition function (useful in model selection).\footnote{From now on, we remove the dependence on ${\bf y}$ in order to simplify the notation.}
The goal is to compute some moment of $\x$, i.e., an integral measure w.r.t. the target pdf,
\begin{equation}
	I = \frac{1}{Z} \int f(\x) \pi(\x) d\x,
\label{eq:integral}
\end{equation}
where $f$ can be any square integrable function of $\x$ w.r.t. $\pi(\x)$, and $Z = \int \pi(\x) d\x$.{\footnote{{Let us recall that $f(\x)$ is square integrable w.r.t. $\pi(\x)$ if $f(\x) \in L_{\pi}^2$, i.e., if $\int_{\mathcal{X}}{f(\x)^2 \pi(\x) d\x} < \infty$.}}}

In many practical applications, both the integral \eqref{eq:integral} and $Z$ cannot be obtained in closed form and must be approximated. Importance sampling methods allow for the approximation of both quantities by a set of properly weighted samples.

\section{Population Monte Carlo (PMC)}
\label{sec_standard_PMC}

\subsection{Description of the original PMC algorithm}
\label{subsec-pmc_method}

The PMC method \cite{Cappe04} is a well-known iterative adaptive importance sampling technique.
At each iteration it generates a set of $N$ samples $\{\x_i^{(t)}\}_{i=1}^N$, where $t$ denotes the iteration number and $i$ denotes the sample index.
In order to obtain the samples, the original PMC algorithm makes use of a collection of proposal densities $\{q_i^{(t)}(\x)\}_{i=1}^N$, with each sample being drawn from a different proposal, $\x_i^{(t)}\sim q_i^{(t)}(\x)$ for $i = 1,\ldots, N$.
Then, they are assigned an importance weight, computed as $w_i^{(t)}=\frac{\pi(\x_i^{(t)})}{q_i^{(t)}(\x_i^{(t)})}$, i.e., the weight of a particular sample represents the ratio between the evaluation, at the sample value, of the target distribution and the evaluation at the sample value of the proposal used to generate it.
The method proceeds iteratively (up to the maximum iteration step  considered, $T$), building a global importance sampling estimator using different proposals at every iteration.
The new proposals are obtained by updating the set of proposals in the previous iteration. 

There are two key issues in the application of PMC methods: the adaptation of the proposals from iteration to iteration and the way resampling is applied.
The latter is critical to avoid the degeneracy of the random measure, i.e., to avoid a few particles having extremely large weights and the rest negligible ones \cite{Robert04,Liu04b}.
Through the resampling procedure one selects the most promising streams of samples from the first iteration up to the current one.
Several resampling procedures have been proposed in the literature \cite{Douc05,delMoral2012adaptive}.
In the standard PMC \cite{Cappe04}, multinomial resampling is the method of choice, and consists of sampling $N$ times from the discrete probability mass defined by the normalized weights.
As a result of this procedure, the new set of parameters used to adapt the proposals for the generation of samples in the next iteration is selected.
In summary, the standard PMC technique consists of the steps shown in Table \ref{PMCalg}.

\begin{table}[!t]
	\centering
	\caption{\textbf{Standard PMC algorithm \cite{Cappe04}.}}
	\begin{tabular}{|p{0.95\columnwidth}|}
    \hline
		\footnotesize
		\begin{enumerate}
			\item {\bf [Initialization]}: Select the parameters defining the $N$ proposals:
				\begin{itemize}
					\item The adaptive parameters ${\bf \mathcal{P}}^{(1)}=\{{\bm \mu}_1^{(1)},...,{\bm \mu}_N^{(1)}\}$.
					\item The set of static parameters, $\{\bC_i \}_{i = 1}^N$.
				\end{itemize}
			E.g., if the proposals were Gaussian distributions one could select the adapting parameters in ${\bf \mathcal{P}}^{(1)}$ as the means of the proposals (that would be updated through the iterations) and the static parameters $\{\bC_i \}_{i = 1}^N$ as their covariances \cite{Cappe04}.
			\vspace*{6pt}
			\item {\bf[For $\bm t \bm= \bm 1$ to  $\bm T$]}: 
			\begin{enumerate}
				\item Draw one sample from each proposal pdf,
					\begin{equation} 
						\x_i^{(t)} \sim q_i^{(t)}(\x|\bmu_i^{(t)},\bC_i), \quad \quad  i=1,\ldots,N.
						\label{eq_drawing_std_pmc}
					\end{equation}
				\item Compute the importance weights,
					\begin{equation} 
						w_i^{(t)}=\frac{\pi(\x_i^{(t)})}{q_i^{(t)}(\x_i^{(t)}|\bmu_i^{(t)},\bC_i)}, \quad \quad  i=1,\ldots,N,
					\label{is_weights}
					\end{equation}
					and normalize them,
					\begin{equation} 
						\bar{w}_i^{(t)}=\frac{w_i^{(t)}}{\sum_{j=1}^N w_j^{(t)}}, \quad \quad  i=1,\ldots,N.
					\label{resampling_weights}
					\end{equation}
				\item Perform multinomial resampling by drawing $N$ independent parameters ${\bm \mu}_i^{(t+1)}$ from the discrete probability {random measure},
					\begin{equation}
						\hat{\pi}^{{N}}_{{t}}(\x)=\sum_{i=1}^N \bar{w}_i^{(t)} \delta(\x-\x_i^{(t)}).
					\label{eq:aproxPi}
					\end{equation}
					The new set of adaptive parameters defining the next population of proposals becomes					\begin{equation}
						\label{resampled_pop}
						{\bf \mathcal{P}}^{(t+1)} =\{{\bm \mu}_1^{(t+1)},...,{\bm \mu}_N^{(t+1)}\}. 
					\end{equation}
			\end{enumerate}
			\item {\bf [Output, $\bm t \bm =\bm T$]}: 
				Return the pairs $\{\x_i^{(t)}, \bar{\rho}_i^{(t)}\}$, with $\bar{\rho}_i^{(t)}$ given by Eq. \eqref{Rho1}, for
				$i=1,\ldots,N$ and $t=1,\ldots,T$.
		\end{enumerate} \\
		\hline 
\end{tabular}
\label{PMCalg}
\end{table}

\subsection{Estimators and consistency}
\label{subsec-estimators}
{\color{black}
All the generated samples can be used to build a global approximation of the target.
This can be done by first normalizing all the weights from all the iterations, 
\begin{equation} 
\label{Rho1}
\bar{\rho}_i^{(t)}=\frac{w_i^{(t)}}{\sum_{\tau=1}^{t}\sum_{j=1}^{N} w_j^{(\tau)}}, \quad \quad t=1,\ldots,T, \quad i=1,\ldots,N,
\end{equation}
and then providing the pairs $\{\x_i^{(t)}, \bar{\rho}_i^{(t)}\}$ for $i=1,\ldots,N$ and $t=1,\ldots,T$.
This procedure to compute the weights is equivalent to applying a static importance sampling technique that considers $NT$ different proposal pdfs and all the corresponding samples. If the normalizing constant $Z$ is known,  the integral in Eq. \eqref{eq:integral} is approximated by the unbiased estimator 
\begin{equation}
	\hat{I}_t =  \frac{1}{tN} \frac{1}{Z} \sum_{\tau=1}^t \sum_{j=1}^N w_j^{(\tau)} f(\x_j^{(\tau)}).
\label{eq_partial_estimator_unnorm}
\end{equation}
When the normalizing constant is unknown, the unbiased estimate of $Z$ is substituted in Eq. \eqref{eq_partial_estimator_unnorm}, yielding the self-normalized estimator
\begin{equation}
	\tilde{I}_t = \sum_{\tau=1}^t \sum_{j=1}^N \bar{\rho}_j^{(\tau)} f(\x_j^{(\tau)})=\frac{1}{tN} \frac{1}{\hat{Z_t}} \sum_{\tau=1}^t \sum_{j=1}^N w_j^{(\tau)} f(\x_j^{(\tau)}) ,
\label{eq_partial_estimator_self_norm}
\end{equation}
where
\begin{equation}
	\hat{Z}_t = \frac{1}{tN}  \sum_{\tau=1}^t \sum_{j=1}^N  w_j^{(\tau)},
\label{eq_Z_estimator}
\end{equation}
is the unbiased estimate of the normalizing constant.}

\section{Improved PMC schemes}
\label{sec_extensions}
{\color{black} 
In the following, we introduce several alternative strategies that decrease the variance of the estimators by exploiting the mixture perspective, and improve the diversity of the population w.r.t. to the standard PMC.
More specifically, we study three different PMC schemes: one related to the strategy for calculating the weights and the other two based on modifying the way in which the resampling step is performed.
Although we concentrate on the standard PMC, we remark that these alternative schemes can be directly applied or combined in other more sophisticated PMC algorithms. 
{Moreover, the alternative schemes can be easily implemented in other Monte Carlo methods with resampling steps, such as the Sequential Monte Carlo (SMC) samplers \cite{del2006sequential}, as we show in \hbox{Sections \ref{sec_toy_example} and \ref{sec_high_dimension}}.}
}

\subsection{Scheme 1:Deterministic mixture PMC (DM-PMC)}
\label{section_dm}

{The underlying idea of PMC is to perform a good adaptation of the location parameters ${\bm \mu}_{i}^{(t)}$, i.e., where the proposals of the next iteration will be centered (e.g., if $q_i^{(t)}$ is a Gaussian pdf, then ${\bm \mu}_{i}^{(t)}$ is its mean). 
These parameters} are obtained at each iteration by sampling from $\cblack{\hat{\pi}_{t-1}^{N}}$ in Eq. \eqref{eq:aproxPi} (i.e., via resampling), which is a random measure that approximates the target distribution, i.e., ${\bm \mu}_{i}^{(t)}\sim \hat{\pi}_{t-1}^{N}$.
As a direct consequence of the strong law of large numbers, $\hat{I}_t \to I$ almost surely (a.s.) as $N \to \infty$ under very weak assumptions \cite{Geweke1989} (the support of the proposal includes the support of the target and $I < \infty$).
{Furthermore, by setting $f_{\z}(\X) = \mathbb{I}(\X \le \z)$, where $\X = [X_1, \ldots, X_{D_x}]$, $\z = [z_1, \ldots, z_{D_x}]$, and $\mathbb{I}(\X \le \z)$ is defined as
\begin{equation*}
	\mathbb{I}(\X \le \z) = \prod_{d=1}^{D_x}{\mathbb{I}(X_d \le z_d)},
\end{equation*}
where $\mathbb{I}(X_d \le z_d)$ denotes the indicator function for the $d$-th component ($1 \le d \le D_x$) of the variable of interest,
\begin{equation*}
	\mathbb{I}(X_d \le z_d) =
		\begin{cases}
			1, & X_d \le z_d;\\
			0, & X_d > z_d,
		\end{cases}
\end{equation*}
then $I = I(\z)$ becomes the multi-variate cumulative distribution function (cdf) of $\pi(\z)$.
Consequently, since $\hat{I}_t(\z) \to I(\z)$ a.s. for any value of $\z$ as $N \to \infty$ {[Geweke,1989]}, ${\bm \mu}_{i}^{(t)} \sim \pi(\x)$ a.s. as $N \to \infty$.
In short, since the cdf associated to $\hat{\pi}_{t-1}^{N}(\x)$ (which is the pdf used for resampling) converges to the target cdf (i.e., the cdf associated to $\pi(\x)$) as $N \to \infty$, then the outputs of the resampling stage (i.e., the means ${\bm \mu}_{i}^{(t)}$) are asymptotically distributed as the target.}

Therefore, the equally-weighted mixture of the set of proposals at the $t$-th iteration, given by
\begin{equation}
\label{completeProposal}
  \psi^{(t)}(\x)= \frac{1}{N}\sum_{i=1}^N q_i^{(t)}(\x|\bmu_i^{(t)},\bC_i),
\end{equation}
can be seen as a kernel density approximation of the target pdf, where the proposals, $\{q_i^{(t)}(\x|\bmu_i^{(t)},\bC_i)\}_{i=1}^N$, play the role of the kernels \cite[Chapter 6]{scott2009multivariate}.
{In general, this estimator has non-zero bias and variance, depending on the choice of $q$, ${\bf C}_i$, and the number of samples, $N$.
However, for a given value of $N$, there exists an optimal choice of ${\bf C}_i^*$ which provides the minimum Mean Integrated Square Error (MISE) estimator \cite{Wand94}.
Using this optimal covariance matrix ${\bf C}_i^*$, it can be proved that
 \begin{equation}
 \label{Esto_CazzVic2}
  \psi^{(t)}(\x)= \frac{1}{N}\sum_{i=1}^N q_i^{(t)}(\x|\bmu_i^{(t)},\bC_i^*) \rightarrow \widetilde{\pi}({\bf x})
\end{equation}
pointwise as $N \rightarrow \infty$ \cite{Wand94}.
Hence, resampling naturally leads to a concentration of the proposals around the modes of the target for large values of $N$.}
 
Therefore, since the performance of an importance sampling method relies on the discrepancy between the numerator (the target) and the denominator (usually, the proposal pdf), a reasonable choice for calculating the importance weights is
\begin{equation} 
	w_i^{(t)}=\frac{\pi(\x_i^{(t)})}{\psi^{(t)}(\x_i^{(t)})}=\frac{\pi(\x_i^{(t)})}{\frac{1}{N}\sum_{j=1}^{N}q_j^{(t)}(\x_i^{(t)}|\bmu_j^{(t)},\bC_j)},
\label{dm_weights}
\end{equation}
where, as opposed to Eq. \eqref{is_weights}, the complete mixture of proposals $\psi(\x)$ is accounted for in the denominator. 

\subsubsection{Theoretical justification}

The first justification for using these \emph{deterministic mixture} (DM) weights is merely mathematical, since the estimator $\hat I_t$ of Eq. \eqref{eq_partial_estimator_unnorm} with these weights is also unbiased (see the proof in \ref{DM_unbiased_appendix}).
The main advantage of this new scheme is that it yields more efficient estimators, i.e. with less variance, combining the deterministic mixture sampling (as in standard PMC) with the weight calculation that accounts for the whole mixture. 
Namely, the estimator ${\hat I}_t$ in Eq. \eqref{eq_partial_estimator_unnorm}, computed using the DM approach, has less variance than the estimator obtained by the standard PMC, as proved in \ref{DM_variance_appendix} for any target and set of proposal pdfs. {These DM weights have been explored in the literature of multiple importance sampling (see for instance the balance heuristic strategy of \cite[Section 3.3.]{Veach95} or the deterministic mixture approach of \cite[Section 4.3.]{Owen00}).} 

The intuition behind the variance reduction is clear in a multi-modal scenario, where different proposals have been successfully adapted covering the different modes, and therefore, the whole mixture of proposals has less mismatch w.r.t. the target than each proposal separately. Indeed, it can be easily proved that the mismatch of the whole mixture w.r.t to the target is always less than the average mismatch of each proposal. {More precisely, let us consider the $L_p$ functional distance (with $p > 1$) among the target and an arbitrary function $g(\x)$,
\begin{equation}
	D_p(\normalized\pi(\x),g(\x)) = \left[\int |\normalized \pi(\x) - g(\x)|^p d\x\right]^{1/p},
\label{eq:lp_dist}
\end{equation}
and let us recall Jensen's inequality \cite{hardy1952inequalities},
\begin{equation}
	\varphi \left(\sum_{i=1}^{N}{\alpha_i z_i}\right) \le \sum_{i=1}^{N}{\alpha_i \varphi \left(z_i\right)},
\label{eq:jensen}
\end{equation}
which is valid for any convex function $\varphi(\cdot)$, any set of non-negative weights $\alpha_i$ such that $\sum_{i=1}^{N}{\alpha_i}=1$, and any collection of points $\{z_i\}_{i=1}^{N}$ in the support of $\varphi$.
Then, by using Jensen's inequality in Eq. \eqref{eq:jensen} with $\varphi(z(\x)) = \left[\int |z(\x)|^p d\x\right]^{1/p}$, $\alpha_i = \frac{1}{N}$ and $z_i = \normalized \pi(\x) - q_i(\x)$, it is straightforward to show that
\begin{equation}
	D_p(\normalized\pi(\x),\psi(\x)) = D_p\left(\normalized\pi(\x),\frac{1}{N} \sum_{i=1}^{N}{q_i(\x)}\right)
		\le \frac{1}{N} \sum_{i=1}^{N}{D_p(\normalized\pi(\x),q_i(\x))}.
\end{equation}
Indeed, although we have focused on the $L_p$ distance, the proof is valid for any distance function which is based on a norm (i.e., any distance s.t. $\varphi(z(\x)) = \|z(\x)\|$ for some norm $\|\cdot\|$), {since every norm is a convex function.}}

Another benefit of the DM-PMC scheme is the improvement in the exploratory behavior of the algorithm.
Namely, since the weights in DM-PMC take into account all the proposals (i.e., the complete mixture) for their calculation, they temper the overrepresentation of high probability areas of the target.
{Note that, as a consequence of the variance reduction of the DM weights, the effective sample size in DM-PMC in a specific iteration is larger than with the standard IS weights.\bfootnote{The effective sample size is the number of independent samples drawn from the target distribution that are equivalent (in terms of variance of the estimators) to the performance of the $N$ samples used in the importance sampling estimator.} The expression $\widehat{ESS} = \frac{1}{\sum_{n}^N \bar{w}_n^2}$ is widely used as a sample approximation of the effective sample size (see its derivation in \cite{kong1992note}). Therefore, if the true underlying effective sample size (the ratio of variances) is larger with the DM weights (than with the standard IS weights), a similar behavior can be considered for $\widehat{ESS}$. As a consequence, the diversity loss associated to the resampling step is reduced by using with the DM weights.
See \cite{elvira2015generalized} for a more detailed discussion about effective sample size in static multiple importance sampling schemes.
}

\subsubsection{Computational complexity discussion}

{In this DM-PMC scheme, the performance is improved at the expense of an increase in the computational cost {(in terms of proposal evaluations)} in the calculation of the weights. 
 However, it is crucial to note that all the proposed schemes keep the same number of evaluations of the target as in the standard PMC. 
 Hence, if the target evaluation is much more costly than the evaluation of the proposal pdfs (as it often happens in practical applications), the increase in computational cost can be negligible in many scenarios of interest.
Note that other adaptive multiple IS algorithms, e.g. \cite{Douc07b,Cappe08,iacobucci2010variance}, also increase the number of proposal evaluations, and they state that the most significant computational cost is associated to the evaluation of the target (see this argument in \cite[Section 2.2.]{Cappe08}).}

{Finally, note that the variant \emph{partial}-DM proposed in \cite{elvira2015efficient} within the static multiple IS framework, could be easily adapted to the DM-PMC. In this weighting scheme, a partition (forming subsets) of the set of proposals is a priori performed. The weight of each sample only accounts at the denominator for a subset of proposals, i.e., reducing the number of proposal evaluations. This variant achieves an intermediate point in the complexity-performance tradeoff, between the standard weights and the DM weights.}

\subsubsection{Comparison with other methods}

{Note that other methods also use a mixture of proposals at the denominator of the weights. For instance, in the D-kernel of \cite{Douc07b}, each sample is drawn from a mixture of $D$ kernels (proposals), and this same mixture is evaluated at the denominator of each weight. Nevertheless, note that these $D$ kernels are centered at the same position, and the weight of each sample ignores the locations of the $N-1$ proposals. In the M-PMC of \cite{Cappe08}, a single mixture is used for sampling and weighting the $N$ samples at each iteration. Note that this method does not use an explicit resampling step, and the mixture is completely adapted  (weights, means, and covariances).}

In the sequel, we adopt the weights of \cblack{Eq. \eqref{dm_weights}} for the other two proposed PMC schemes due to their theoretical and practical advantages discussed above.
\subsection{Scheme 2: Multiple samples per mixand with global resampling (GR-PMC)}
\label{sec_gr}

We propose to draw $K$ samples per individual proposal or mixand, instead of only one as done in the standard PMC algorithm.
Namely,
\begin{equation} 
\x_{i,k}^{(t)} \sim q_i(\x|\bmu_i^{(t)},\bC_i)
\end{equation}
for $i=1,\ldots,N$ and $k=1,\ldots,K$.
Then, we compute the corresponding DM weights as in \eqref{dm_weights},
\begin{equation} 
w_{i,k}^{(t)}=\frac{\pi(\x_{i,k}^{(t)})}{\frac{1}{N}\sum_{j=1}^{N}q_{j}^{(t)}(\x_{i,k}^{(t)}|\bmu_j^{(t)},\bC_j)}.
\label{dm_k_weights}
\end{equation}
Therefore, at each iteration we have a set of $KN$ generated samples, i.e., 
$
{\bf \mathcal{X}}^{(t)}=\{\x_{1,1}^{(t)},...,\x_{1,K}^{(t)},...,\x_{N,1}^{(t)},...,\x_{N,K}^{(t)}\}.
$
Resampling is performed in the same way as in standard PMC, although now the objective is to downsample, from $KN$ samples to $N$ samples, according to the normalized weights,
\begin{equation}
	\bar{w}_{i,k}^{(t)}=\frac{w_{i,k}^{(t)}}{\sum_{j=1}^{N}\sum_{\ell=1}^{K} w_{j,\ell}^{(t)}}.
\label{resampling_weights_gr}
\end{equation}

We refer to this type of resampling as {\it global} resampling, since all the samples, regardless of the proposal used to generate them, are resampled together.
After resampling, a new set of adapted parameters for the next iteration, ${\bf \mathcal{P}}^{(t+1)}=\{{\bm \mu}_1^{(t+1)},...,{\bm \mu}_N^{(t+1)}\}$, is obtained. {Note that, through this paper, for sake of simplicity in the explanation of the proposed improvements, we use the standard multinomial resampling, but other resampling schemes that reduce the path-degeneracy problem can be considered instead, e.g. the residual or stratified resampling, (see \cite{Cappe05,Douc05}).}

{The PMC algorithms suffer from sample impoverishment, which is a side effect inherent to adaptive algorithms with resampling steps such as SMC samplers or particle filters (see for instance \cite[Section V-C]{arulampalam2002tutorial} or \cite[Section 2]{li2014fight}). In other words, there is a diversity reduction of the samples after the resampling step (in a very adverse scenario, the $N$ resampled samples can be $N$ copies of the sample). The sample impoverishment of the standard PMC is illustrated in Fig. \ref{figTrellis}, Fig. \ref{fig_survival}, and Fig. \ref{fig_positions_evolutions}, where the increase of diversity of the algorithms proposed in this paper is shown by numerical simulations. These figures correspond to the example of Section \ref{sec_toy_example} and will be properly introduced below.}
In multimodal scenarios, proposals of the standard PMC that are exploring areas with negligible probability masses are very likely to be removed before they find unexplored relevant areas.
{If we draw} $K$ samples per proposal, the samples of a well-placed proposal will have similarly high weights, but as for the explorative proposals, increasing $K$ also increases their chances of discovering local relevant features of the target $\normalized{\pi}(\x)$.
Then, the GR-PMC promotes the local exploration of the explorative proposals, increasing the chances of not being removed in the resampling step. Figures \ref{figTrellis} and \ref{fig_survival} show the reduction of path-degeneracy of GR-PMC in a multimodal scenario, and they will be properly explained in the example of Section \ref{sec_toy_example}.

{Note that using $K>1$ does not entail an increase in the computational cost w.r.t. the standard PMC or DM-PMC (where $K=1$) if the number of evaluations of the target is fixed to $L=KNT$.
Indeed, since the number of resampling stages is reduced to $T=L/(KN)$, the computational cost decreases, although at the expense of performing less adaptation steps than for $K=1$.
Therefore, for a fixed budget of target evaluations $L$ and a fixed number of proposals $N$, one must decide whether to promote the local exploration (possibly reducing the path degeneracy) by increasing $K$, or performing more adaptation steps $T$.
Thus, there is a trade-off between local and global exploration as the numerical experiments will also show in Section \ref{sec_results}.
This suggests that, for a fixed computational budget $L$, there exists an optimal value of samples per proposal and iteration, $K^*$, which will also depend on the target and cannot be found analytically.
This issue can be partially addressed through the use of local resampling, as shown in the following section.}

\subsection{Scheme 3: Multiple samples per mixand with local resampling (LR-PMC)}
\label{LRsect}

\begin{figure*}[!htb]
\centering
\begin{tikzpicture}
	[box/.style={rectangle, draw=black, fill=gray!30},scale=0.75, transform shape,
	 label/.style={rectangle}]
	\node (inf) at (0,0) [box] {$x_{N,1}^{(t)},\ \ldots,\ x_{N,k}^{(t)},\ \ldots,\ x_{N,K}^{(t)}$};
	\node (med) at (0,2) [box] {$x_{i,1}^{(t)},\ \ldots,\ x_{i,k}^{(t)},\ \ldots,\ x_{i,K}^{(t)}$};
	\node (sup) at (0,4) [box] {$x_{1,1}^{(t)},\ \ldots,\ x_{1,k}^{(t)},\ \ldots,\ x_{1,K}^{(t)}$};
	\node (local) at (5,2) [label] {Local Resampling};
	\node (qN) at (-4,0) [label] {$q_N^{(t)}(\x)$};
	\node (qi) at (-4,2) [label] {$q_i^{(t)}(\x)$};
	\node (q1) at (-4,4) [label] {$q_1^{(t)}(\x)$};
	\node (global) at (0,5) [label] {Global Resampling};
	\draw[>=triangle 45, ->,line width=1.4pt] (local.north west) -- (sup.east);
	\draw[>=triangle 45, ->,line width=1.4pt] (local.west) -- (med.east);
	\draw[>=triangle 45, ->,line width=1.4pt] (local.south west) -- (inf.east);
	\draw[>=triangle 45, ->,line width=1.4pt] (q1.east) -- (sup.west);
	\draw[>=triangle 45, ->,line width=1.4pt] (qi.east) -- (med.west);
	\draw[>=triangle 45, ->,line width=1.4pt] (qN.east) -- (inf.west);
	\draw (-2.7,-0.7) rectangle (2.7,4.7);
\end{tikzpicture}
\caption{Sketch of the global and local resampling schemes considering $N$ proposal pdfs at the $t$-th iteration, $q_i^{(t)}$ for $i=1,\ldots,N$ and $t=1,\ldots,T$, and $K$ samples per proposal.}
\label{fig_GlobLoc}
\end{figure*}
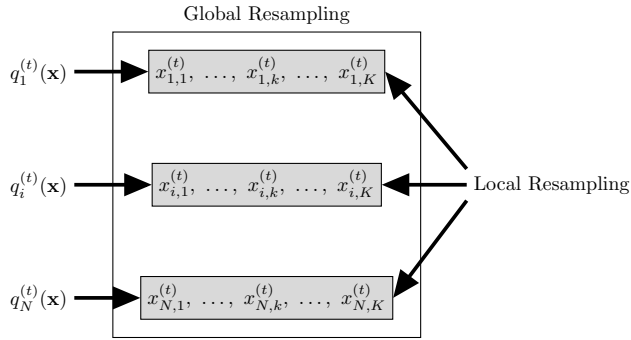

Consider again $K$ samples generated from each proposal pdf.
{In this alternative scheme, the estimators are built as in GR-PMC, i.e., with the weights of Eq. \eqref{dm_k_weights}.
Nevertheless, unlike the previous method, here the resampling step is performed independently for each proposal.}
Namely, at the $t$-th iteration, $K$ samples are drawn from each of the $N$ proposal pdfs, and $N$ {\it parallel} resampling procedures are independently performed within each subset of $K$ samples (see Fig. \ref{fig_GlobLoc} for a visual comparison of both resampling schemes).
More precisely, the adaptive parameter for the next iteration of the $i$-th proposal, ${\bm \mu}_i^{(t+1)}$ for $i=1,\ldots,N$, is resampled from the set 
\begin{equation} 
{\bf \mathcal{X}}_i^{(t)}=\{\x_{i,1}^{(t)},...,\x_{i,K}^{(t)}\},
\end{equation}
using the multinomial probability mass function with probabilities
\begin{equation} 
\bar{w}_{i,k}^{(t)}=\frac{w_{i,k}^{(t)}}{\sum_{\ell=1}^{N} w_{i,\ell}^{(t)}}, \quad \quad k=1,\ldots,K. 
\label{resampling_weights_lr}
\end{equation}
{where the unnormalized weights $w_{i,k}^{(t)}$ are given by Eq. \eqref{dm_k_weights}. 
Note that again we can use any resampling technique, including the standard multinomial or other advanced schemes \cite{Cappe05,Douc05}.
In LR-PMC, there is no loss of diversity in the population of proposals, since each proposal at the current iteration yields another proposal in the next iteration. In other words, exactly one particle per proposal survives after the resampling step.}

{The adaptation scheme of LR-PMC can be intuitively understood as follows. 
Let us consider for a moment a modified version of LR-PMC where the weights used in the resampling are those of standard PMC of Eq. \eqref{resampling_weights} instead of the DM weights of Eq. \eqref{resampling_weights_lr}. 
This modified scheme is equivalent to $N$ parallel PMC samplers, where the $i$-th PMC draws $K$ samples from the $i$-th proposal, applying a resampling step independently from the other $N-1$ PMC samplers. 
By using the DM weights in LR-PMC, we incorporate cooperation among the $N$ proposals. 
When the proposal pdfs are close to each others, the local resampling scheme (with DM weights) adds a ``repulsive'' interaction: among the $K$ samples of a specific proposal, the resampling promotes the samples in areas that are less covered by the other $N-1$ proposals (and where, at the same time, the target evaluation is high). 
Therefore, this scheme performs a cooperative exploration of the state space by the $N$ proposals.
Note that, when the proposal pdfs are located far away from each other, the weights of the $K$ samples of a specific proposal are in practice not affected by other $N-1$ proposals. 
In this case, the LR-PMC works as the $N$ parallel PMC samplers described above.}

Finally, let us remark that a mixed global-local resampling strategy (e.g., performing local resampling on clusters of proposals) could also be devised in order to obtain the advantages of both global and local resampling.

\section{Numerical results}
\label{sec_results}

\subsection{\cblack{Estimation of the normalizing constant}}
\label{sec_ex1}
\begin{figure}[htp]
\centering
\subfigure[]{
\includegraphics[width=0.45\textwidth]{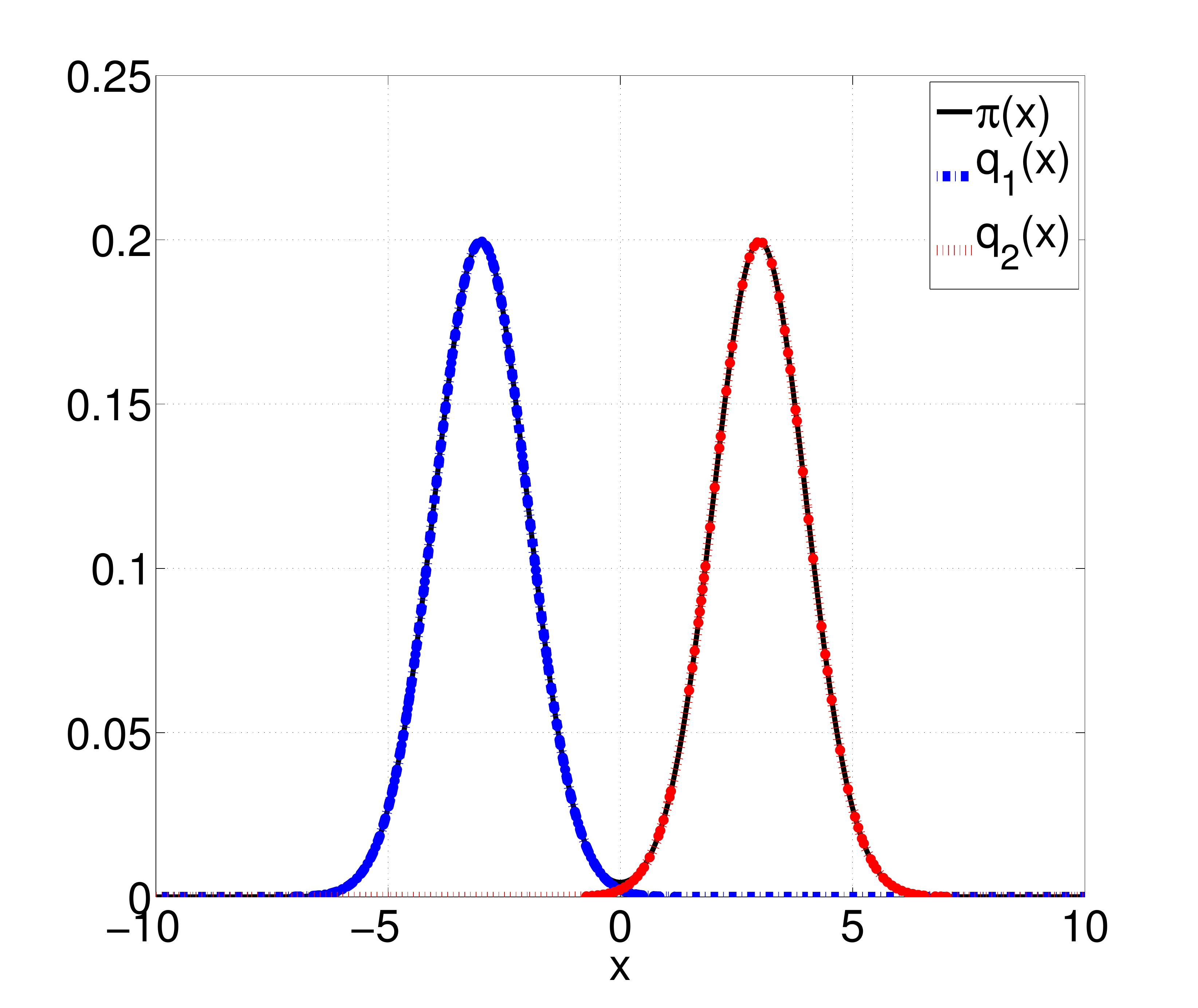}
} 
\centering
\subfigure[]{
\includegraphics[width=0.45\textwidth]{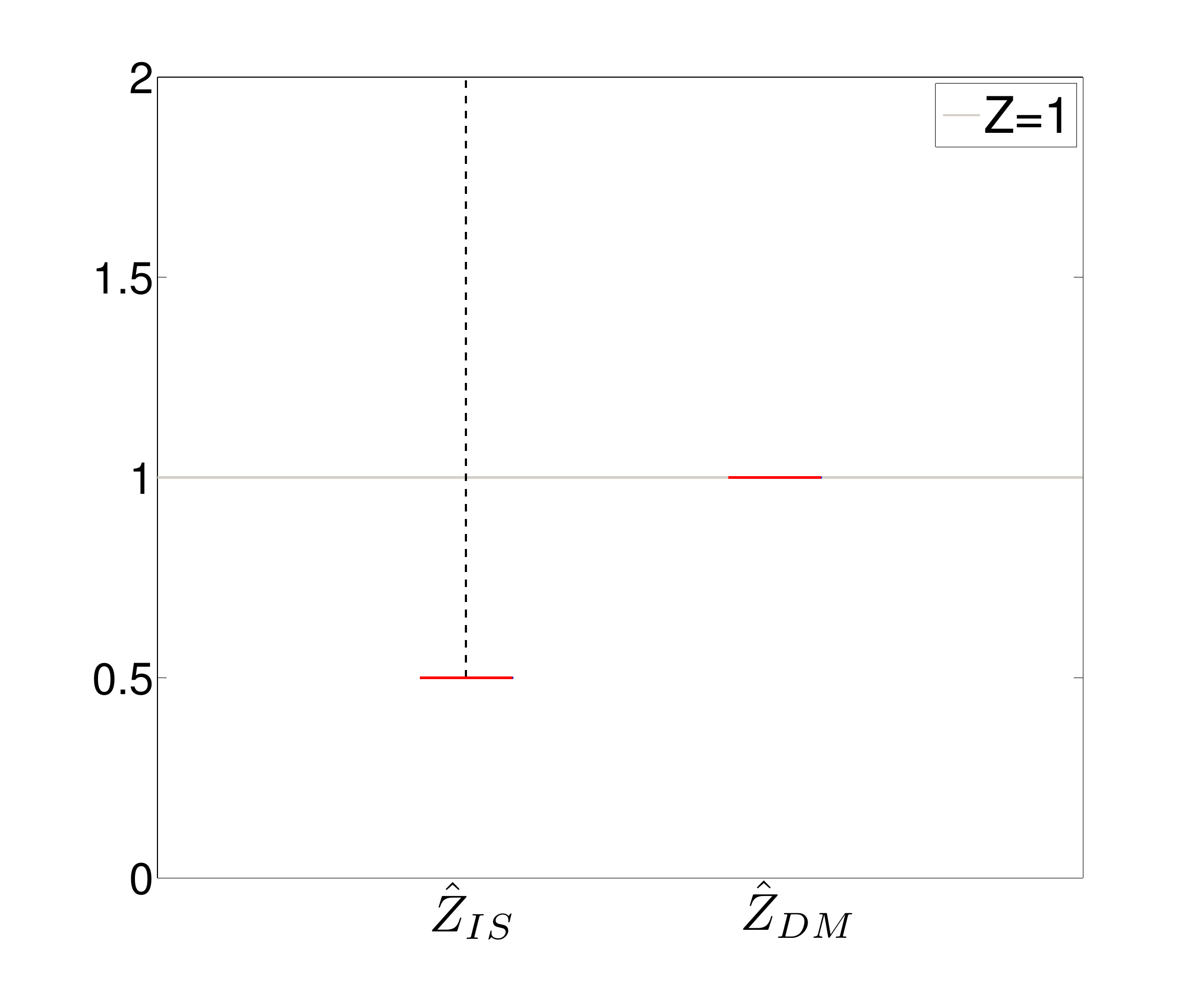}
}
\caption{\textbf{(Ex. of Section \ref{sec_ex1})} Estimation of the normalizing constant (true value $Z=1$) in Scenario 1 (perfect matching). \textbf{(a)} Target pdf (black solid line) and proposal pdfs (red and blue dashed lines). \textbf{(b)} Boxplot showing the 25th and 75th percentiles of the estimators $\hat Z_{IS}$ and $\hat Z_{DM}$. The maximum value of $\hat Z_{IS}$ is $35864$.}
\label{fig_ex_Z_sc1}
\end{figure}

\begin{figure}[htp]
\subfigure[]{
\includegraphics[width=0.45\textwidth]{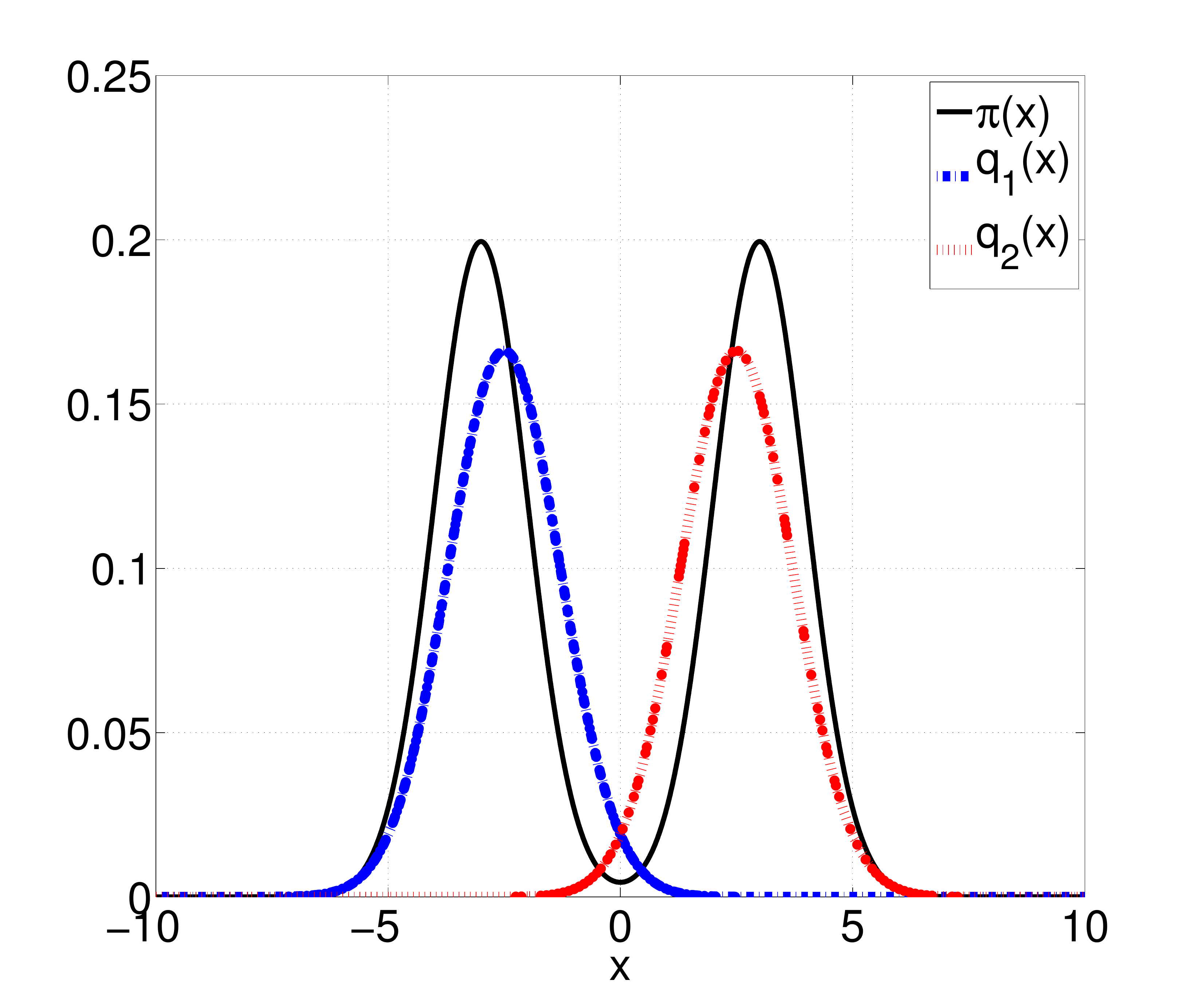}
} 
\centering
\subfigure[]{
\includegraphics[width=0.45\textwidth]{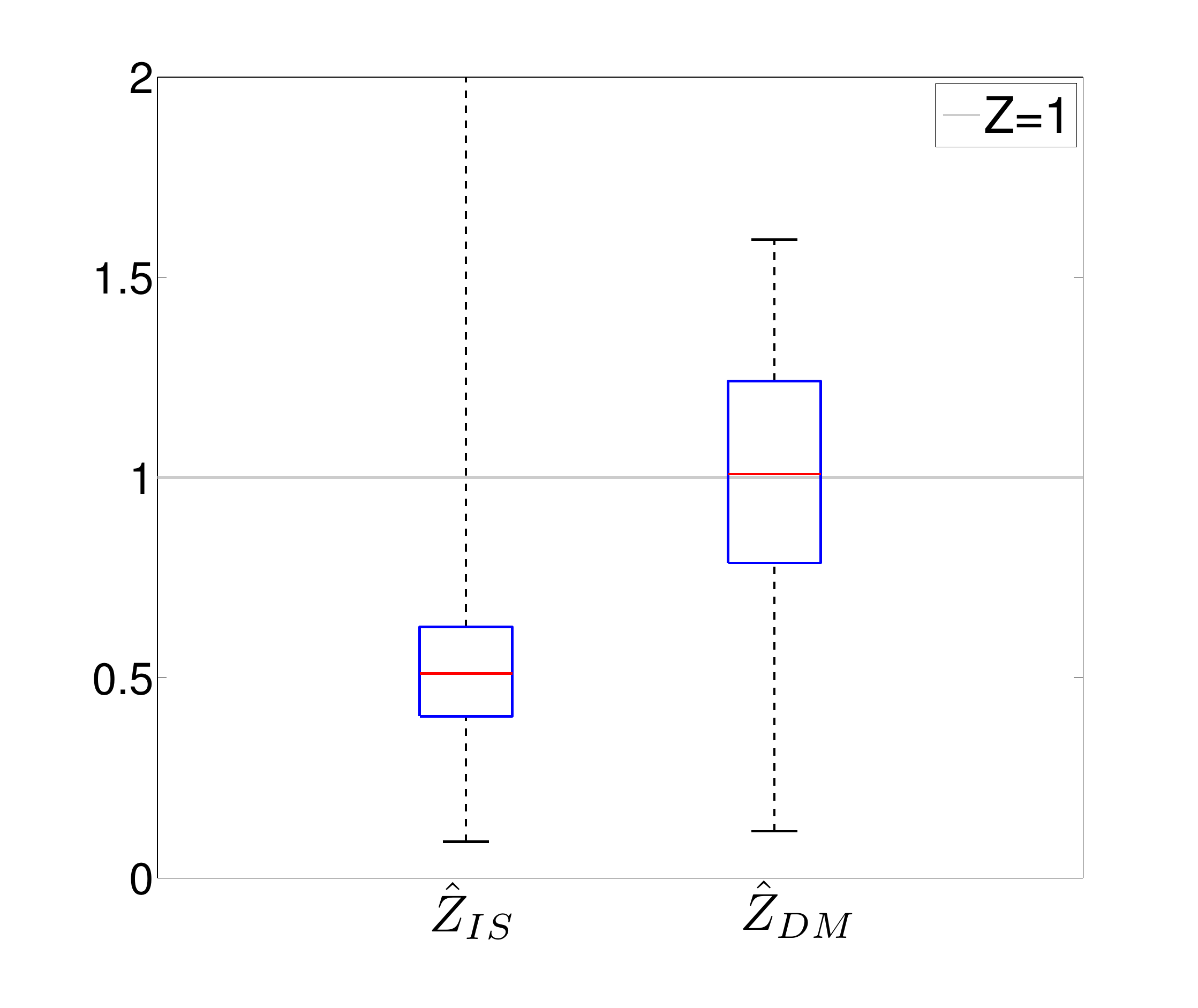}
} 
\caption{\textbf{(Ex. of Section. \ref{sec_ex1})} Estimation of the normalizing constant (true value $Z=1$) in Scenario 2 (proposal-target mismatch). \textbf{(a)} Target pdf (black solid line) and proposal pdfs (red and blue dashed lines). \textbf{(b)} Boxplot showing the distribution of the estimators $\hat Z_{IS}$ and $\hat Z_{DM}$. The maximum value of $\hat Z_{IS}$ is $77238$.}
\label{fig_ex_Z_sc2}
\end{figure}

\cblack{Let us consider, as a target pdf, a bimodal mixture of Gaussians $\pi(\x)=\frac{1}{2} \mathcal{N}(x;\nu_1,c_1^2) + \frac{1}{2} \mathcal{N}(x;\nu_2,c_2^2)$ with $\nu_1=-3$ and $\nu_2=3$, and $c_1^2=1$ and $c_2^2=1$. The proposal pdfs are also Gaussians: $q_1(x)=\mathcal{N}(x;\mu_1,\sigma_1)$ and $q_2(x)=\mathcal{N}(x;\mu_2,\sigma_2)$. At this point, we consider two scenarios:
\begin{itemize}
\item Scenario 1: In this case, $\mu_1 = \nu_1$, $\mu_2 = \nu_2$, $\sigma_1^2 = c_1^2$, and $\sigma_2^2 = c_2^2$. Then, both proposal pdfs can be seen as a whole mixture that exactly replicates the target, i.e., $\pi(\x) = \frac{1}{2} q_1(\x) + \frac{1}{2} q_2(\x)$. This is the desired situation pursued by an adaptive importance sampling algorithm: each proposal is centered at a different mode of the target, and their scale parameters perfectly match the scales of the modes. Fig. \ref{fig_ex_Z_sc1}(a) shows the target pdf in solid black line, and both proposal pdfs in blue and red dashed lines, respectively. Note that the proposals are scaled (each one integrates up to $1/2$ so we can see the perfect matching between the target and the mixture of proposal densities).
\item Scenario 2: In this case, $\mu_1 = -2.5$, $\mu_2 = 2.5$, $\sigma_1^2 = 1.2$, and $\sigma_2^2 = 1.2$. Therefore, there is a mismatch between the target and the two proposals. Fig. \ref{fig_ex_Z_sc2}(a) shows the target pdf in solid black line, and both proposal pdfs in blue and red dashed lines, respectively.
\end{itemize} 
The goal is estimating the normalizing constant using the estimator $\hat Z$ of Eq. \eqref{eq_Z_estimator} with $N=2$ samples, one from each proposal, and $t=1$. We use the standard PMC weights of Eq. \eqref{is_weights} (estimator $\hat Z_{{IS}}$) and the DM-PMC weights of Eq. \eqref{dm_weights} (estimator $\hat Z_{{DM}}$). In order to characterize the two estimators, we run $2 \cdot 10^5$ simulations for each method. Note that the true value is $Z=1$.}

\cblack{Figure \ref{fig_ex_Z_sc1}(b) shows a boxplot of the distribution of the estimator $\hat{Z}$, obtained with both methods for Scenario 1. The blue lower and upper edges of the box correspond to the 25th and 75th percentiles, respectively, while the red line represents the median. The vertical black dashed whiskers extend to the minimum and maximum obtained values. Since the maxima cannot be appreciated in the figure, they are displayed in Table \ref{tabla_maxima}, altogether with the variance of the estimators. Note that even in this extremely simple and idealized scenario (perfect adaptation), the estimator obtained using the standard IS weights (i.e., the estimator used in standard PMC) has a poor performance. In most of the realizations, $\hat Z_{{IS}} \approx 0.5$ because each proposal (which integrates up to one) is adapted to one of the two modes (which contain roughly half of the probability mass).\footnote{\cblack{In this setup, each proposal approximately covers a different half of the target probability mass, since each one coincides with a different mode of the target. However, in standard PMC, the weight of each sample only accounts for its own proposal, and therefore there is not an exchange of information among the two proposals. Note that, if both proposals were covering the same mode (and therefore missing the other one), the weights would also be $w=0.5$ in most of the runs; the lack of information exchange between the two samples, makes it impossible to know whether the target mass reported by the weight of each sample is the same and should be accounted ``once'', or whether it is from another area and it should be accounted ``twice''.}}  Since $E[\hat Z_{{IS}}]=Z=1$, in a few runs the value the $\hat Z_{{IS}}$ is extremely high as shown in Table \ref{tabla_maxima}. These huge values occur when a sample drawn from the tail of the proposal falls close to the other mode of the target (where actually the other proposal is placed).
\cblack{On the other hand, note that the DM estimator has a perfect performance (i.e., $\hat Z_{{DM}}=1$ always, thus implying zero variance).
Hence, this simple example shows that a substantial variance reduction can be attained by using the mixture at the denominator.}}

\cblack{Figure \ref{fig_ex_Z_sc2}(b) shows an equivalent boxplot for Scenario 2. In this case, the mismatch between proposals and target pdfs worsens both schemes. Note that the estimator $\hat Z_{{DM}}$ now does not perfectly approximates $Z$, but still largely outperform the estimator $\hat Z_{{IS}}$. In particular, the median is still around the true value, and its variance is smaller.}
\begin{table*}[!t]
\begin{center}
\begin{tabular}{|c|c|c|c||c|c|c|c|}
\hline
& \cblack{Estimator} & \cblack{$\hat Z_{{IS}}$} & \cblack{$\hat Z_{{DM}}$} &  & \cblack{Estimator} & \cblack{$\hat Z_{{IS}}$} &  \cblack{$\hat Z_{{DM}}$} \\
\hline
\hline
\multirow{2}{*}{\cblack{{\bf (Sc. 1)}}}  & \cblack{Max.} &  \cblack{35864}   &    \cblack{1}  & \multirow{2}{*}{\cblack{{\bf  (Sc. 2)}}}  & \cblack{Max.} & \cblack{77238}  &  \cblack{1.59}\\
\cline{2-4}
\cline{6-8}
& \cblack{$Var(\hat Z)$} &    \cblack{7891}  &   \cblack{0} & &\cblack{$Var(\hat Z)$}  &  \cblack{ 6874}    &   \cblack{0.01} \\
\hline
\end{tabular}
\end{center}
\caption{\cblack{\textbf{(Ex. of Section. \ref{sec_ex1})} Maximum value of the estimator $\hat Z$ in $2\cdot 10^5$ runs for each scheme, in two different scenarios.}}
\label{tabla_maxima}
\end{table*}

\subsection{Bi-dimensional example}
\label{sec_toy_example}

We first consider a bivariate multimodal target pdf, consisting of a mixture of five Gaussians, i.e., 
\begin{equation}
\pi(\x)=\frac{1}{5}\sum_{i=1}^5 \mathcal{N}(\x;{\bm \nu}_i,{\bf \Sigma}_i), \quad \x\in \mathbb{R}^2,
\label{Target1b}
\end{equation}
where $\mathcal{N}(\x;{\bm \mu},{\bf C})$ denotes a normalized Gaussian pdf with mean vector ${\bm \mu}$ and covariance matrix ${\bf C}$, ${\bf \nu}_1=[-10, -10]^{\top}$, ${\bm \nu}_2=[0, 16]^{\top}$, ${\bm \nu}_3=[13, 8]^{\top}$, ${\bm \nu}_4=[-9, 7]^{\top}$, ${\bm \nu}_5=[14, -14]^{\top}$, ${\bf \Sigma}_1=[2, \ 0.6; 0.6, \ 1]$, ${\bf \Sigma}_2=[2, \ -0.4; -0.4, \ 2]$, ${\bf \Sigma}_3=[2, \ 0.8; 0.8, \ 2]$, ${\bf \Sigma}_4=[3, \ 0; 0, \ 0.5]$, and ${\bf \Sigma}_5=[2, \ -0.1; -0.1, \ 2]$. 
In this example, we can analytically compute different moments of the target in \eqref{Target1b}, and therefore we can easily validate the performance of the different techniques.
In particular, we consider the computation of the mean of the target, $E[\X]=[1.6, 1.4]^{\top}$, and the normalizing constant, $Z=1$, for $\x\sim \frac{1}{Z}\pi(\x)$.
We use as figure of merit the Mean Squared Error (MSE) in the estimation of $E[\X]$ (averaged over both components) and $Z$.

\begin{landscape}
\begin{table*} 
\setlength{\tabcolsep}{2pt}
\def\marginwidth{1.5mm}
\begin{center}
{\tiny
\begin{tabular}{|c@{\hspace{\marginwidth}}|l@{\hspace{\marginwidth}}|c@{\hspace{\marginwidth}}|c@{\hspace{\marginwidth}}|c@{\hspace{\marginwidth}}|c@{\hspace{\marginwidth}}|c@{\hspace{\marginwidth}}|c@{\hspace{\marginwidth}}|}
\hline
  \multicolumn{8}{|c|}{ $\bm L= \bm N\bm K\bm T = 2 \cdot 10^5$} \\
\hline
$\bm N$ &{\bf Algorithm} &  ${\bm \sigma{\bf=1}}$ &  ${\bm \sigma{\bf=2}}$ &  ${\bm \sigma{\bf=5}}$ &  ${\bm \sigma{\bf=10}}$ & ${\bm \sigma{\bf=20}}$ & ${\bm \sigma{\bf=70}}$  \\
\hline
\hline
5 & \multirow{ 3}{*}{Standard PMC \cite{Cappe04}} &92.80 (85.23-99.57) &38.71 (31.96-47.69) &12.65 (7.10-19.04) &0.38 (0.28-0.53) &0.047 (0.033-0.065) &37.44 (21.01-55.62)\\
\cline{1-1}
\cline{3-8}
100 & &75.17 (72.72-78.20) &59.42 (54.78-64.23) &14.24 (12.04-16.57) &0.25 (0.21-0.30) &0.028 (0.023-0.033) &0.18 (0.15-0.22)\\
\cline{1-1}
\cline{3-8}
$5 \cdot 10^4$ & &68.29 (66.92-69.19) &37.44 (34.57-41.98) &7.01 (5.72-7.86) &0.25 (0.18-0.34) &0.033 (0.027-0.039) &0.17 (0.14-0.21)\\
\hline
\hline
\multirow{ 15}{*}{} &DM-PMC ($K=1$) & 72.48 (69.79-75.14) &36.21 (33.54-39.26) &5.34 (4.41-6.33) &0.036 (0.030-0.043) &0.029 (0.024-0.034) &0.21 (0.18-0.25)\\
\cline{2-8}
&GR-PMC ($K=2$) &69.41 (66.02-72.30) &26.23 (22.26-30.83) &3.09 (1.88-4.69) &0.022 (0.019-0.027) &0.028 (0.022-0.033) &0.17 (0.14-0.21)\\
&LR-PMC ($K=2$) &\textbf{2.68} (1.85-3.54) & \textbf{0.007} (0.005-0.009) &0.010 (0.008-0.012) &0.018 (0.014-0.022) &0.102 (0.084-0.122) &32.88 (27.89-38.69)\\

\cline{2-8}
&GR-PMC ($K=5$) &67.04 (64.26-69.53) &17.44 (14.74-20.55) &0.11 (0.03-0.25) &0.013 (0.011-0.016) &\textbf{0.023} (0.018-0.027) &0.15 (0.12-0.17)\\
&LR-PMC ($K=5$) &8.04 (6.65-9.65) &0.012 (0.007-0.019) &\textbf{0.008} (0.005-0.012) &0.016 (0.013-0.019) &0.027 (0.021-0.033) &2.00 (1.52-2.60)\\
\cline{2-8}
&GR-PMC ($K=20$) &61.58 (56.94-66.03) &15.13 (12.30-18.81) &0.42 (0.03-1.18) &0.012 (0.010-0.014) &0.024 (0.020-0.029) &\textbf{0.14} (0.12-0.17)\\
&LR-PMC ($K=20$) &9.51 (8.49-10.53) &1.16 (0.54-1.89) &0.011 (0.008-0.014) &0.013 (0.011-0.016) &0.023 (0.019-0.028) &0.22 (0.18-0.26)\\

\cline{2-8}
&GR-PMC ($K=100$) &64.94 (61.67-67.66) &12.50 (10.65-15.53) &0.08 (0.02-0.20) &0.015 (0.011-0.018) &0.026 (0.021-0.030) &0.18 (0.15-0.21)\\
&LR-PMC ($K=100$) &9.60 (8.58-10.66) &1.21 (0.64-1.88) &0.022 (0.016-0.029) &0.015 (0.012-0.018) &0.026 (0.022-0.032) &0.20 (0.16-0.24)\\
\cline{2-8}
&GR-PMC ($K=500$) &58.49 (54.10-62.20) &9.63 (7.81-11.45) &0.08 (0.06-0.10) &0.014 (0.011-0.016) &0.024 (0.019-0.030) &0.16 (0.14-0.20)\\
&LR-PMC ($K=500$) &14.79 (13.12-16.54) &6.72 (5.30-8.39) &0.10 (0.06-0.14) &\textbf{0.010} (0.008-0.013) &0.024 (0.018-0.030) &0.20 (0.16-0.25)\\
\hline 
 100 & \cblack{M-PMC \cite{Cappe08}}& 71.39 (65.22-77.36) &81.33 (71.59-90.04) &18.14 (13.51-22.90) &0.058 (0.052-0.067) &0.031 (0.016-0.056) &\textbf{0.14} (0.11-0.17)\\
\hline
 10 &   &84.14 (73.46-97.81) &81.68 (67.66-95.91) &6.49 (2.58-10.45) &0.76 (0.15-1.71) &\textbf{0.024} (0.021-0.027) &4.60 (1.64-8.51)\\ 
 \cline{1-1}
 \cline{3-8}
  100 & {SMC \cite{del2006sequential}}&77.00 (76.35-77.66) &76.57 (75.60-77.66) &15.98 (15.42-16.59) &0.79 (0.64-0.97) &0.068 (0.065-0.072) &0.86 (0.79-0.93)\\ 
  \cline{1-1}
 \cline{3-8}
$5 \cdot 10^4$  & &69.08 (68.34-69.91) &51.29 (44.10-57.26) &20.48 (8.86-36.70) &0.22 (0.14-0.31) &0.038 (0.019-0.061) &0.68 (0.39-1.03) \\
\hline
   &  {DM-SMC ($K=1$)}  &70.95 (70.16-71.74) &42.40 (41.49-43.39) &1.91 (1.72-2.15) &0.039 (0.037-0.040) &0.027 (0.026-0.029) &0.19 (0.18-0.19)\\
    \cline{2-8}
   &  {GR-SMC ($K=5$)}  &66.64 (65.42-67.84) &41.54 (39.93-43.01) &0.16 (0.15-0.18) &0.015 (0.014-0.016) &\textbf{0.024} (0.023-0.025) &0.19 (0.19-0.20)\\
       \cline{3-8}
    100 &  {LR-SMC ($K=5$)}  &8.16 (7.68-8.66) &2.32 (1.92-2.71) &0.007 (0.006-0.008) &0.015 (0.014-0.016) &0.027 (0.026-0.028) &2.19 (2.08-2.29)\\
        \cline{2-8}
       &  {GR-SMC ($K=20$)} &65.48 (64.16-66.67) &37.91 (36.21-39.75) &0.10 (0.05-0.18) &0.013 (0.012-0.014) &0.025 (0.024-0.026) &0.19 (0.18-0.20)\\
           \cline{3-8}
     &  {LR-SMC ($K=20$)} &8.88 (8.45-9.32) &4.15 (3.65-4.62) &0.010 (0.008-0.012) &0.014 (0.013-0.014) &0.026 (0.025-0.027) &0.20 (0.19-0.20)\\
\hline                                                             
\end{tabular}
}
\end{center}
\caption{{\bf (Ex. of Section \ref{sec_toy_example})} MSE in the estimation of $E[\X]$, for several values of $\sigma$ and $K$, keeping the total number of evaluations of the target fixed to $L=KNT=2 \cdot 10^5$ in all algorithms. The best results for each value of $\sigma$ are highlighted in bold-face.}
\label{table_mean}
\end{table*}
\end{landscape}

For simplicity, we assume Gaussian proposal densities for all of the methods compared, and deliberately choose a ``bad'' initialization of the means in order to test the robustness and the adaptation capabilities.
Specifically, the initial adaptive parameters of the individual proposals are selected uniformly within the $[-4,4]\times[-4,4]$ square, i.e., ${\bm \mu}_i^{(1)}\sim \mathcal{U}([-4,4]\times[-4,4])$ for $i=1,\ldots,N$.
This initialization is considered ``bad'', since none of the modes of the target falls within the initialization square.
We test all the alternatives using the same isotropic covariance matrices for all the Gaussian proposals, $\bC_i = \sigma^2\bI_2$ with $\sigma \in \{1,2,5,10,20,70\}$.
All the results have been averaged over $500$ independent experiments, where the computational cost of the different techniques (in terms of the total number of evaluations of the target distribution) is fixed to $L=KNT$.\footnote{Note that $L=KNT$ also corresponds to the total number of samples generated in all the schemes.}
We compare the following schemes:

\begin{itemize}

\item {\bf Standard PMC \cite{Cappe04}:} Standard PMC algorithm described in Table \ref{PMCalg} with $N=100$ proposals and $T=2000$ iterations. The total number of samples drawn is $L = NT = 2 \cdot 10^5$.

\item \cblack{{\bf M-PMC \cite{Cappe08}:} M-PMC algorithm proposed in \cite{Cappe08} with $D=100$ proposals, $N=100$ samples per iteration, and $T=2000$ iterations. The total number of samples drawn is $L = NT = 2 \cdot 10^5$.}

\item {{\bf SMC \cite{del2006sequential}:}  We apply a Sequential Monte Carlo (SMC) scheme combining resampling and MCMC steps. Specifically, we consider Metropolis-Hastings (MH) steps as forward reversible kernels. In this example, we do not employ a sequence of tempered target pdfs, i.e., we consider always the true target density. 
The proposal pdfs for the MH kernels coincide with the Gaussian proposals employed in the propagation resampling steps, with the scale parameters ${\bf C}_i$ of the other tested methods. Due to the application of the MH steps, in this case, $L>2\cdot 10^5$.    
}

\item {\bf K-PMC:} Standard PMC scheme using $N=100$ proposals, but drawing $K>1$ samples per proposal at each iteration and performing global resampling (GR). In order to keep the total number of samples constant, the number of iterations of the algorithm is now $T = 2 \cdot 10^5 /(KN)$. 

\item {\bf DM-PMC:} Standard PMC using the weights of Eq. \eqref{dm_weights}, $N=100$ proposals, $T=2000$ iterations, and drawing $K=1$ samples per proposal.

\item {\bf GR-PMC:} PMC scheme with multiple samples per mixand ($K$), weights computed as Eq. \eqref{dm_k_weights}, and global resampling (GR). We use $N=100$ proposals and $T=L/(KN)$ iterations with $L=2 \cdot 10^5$ (as in the three previous schemes). In particular, we test the values $K \in \{2,5,20,100,500\}$, and thus $T \in \{1000,400,100,20,4\}$.

\item {\bf LR-PMC:} PMC scheme with multiple samples per mixand ($K$) and local resampling (LR). All the parameters are selected as in the GR-PMC scheme.

\item {\textbf{Improved  SMC:} SMC scheme with the improvements proposed in this paper. In all cases, we use the weights of Eq. \eqref{dm_weights} (DM-SMC), and we try the GR-SMC and LR-SMC variants. We test $K \in \{5,20\}$ }

\end{itemize}

\begin{figure*}[!htb]
\centering
\includegraphics[width=0.99\textwidth]{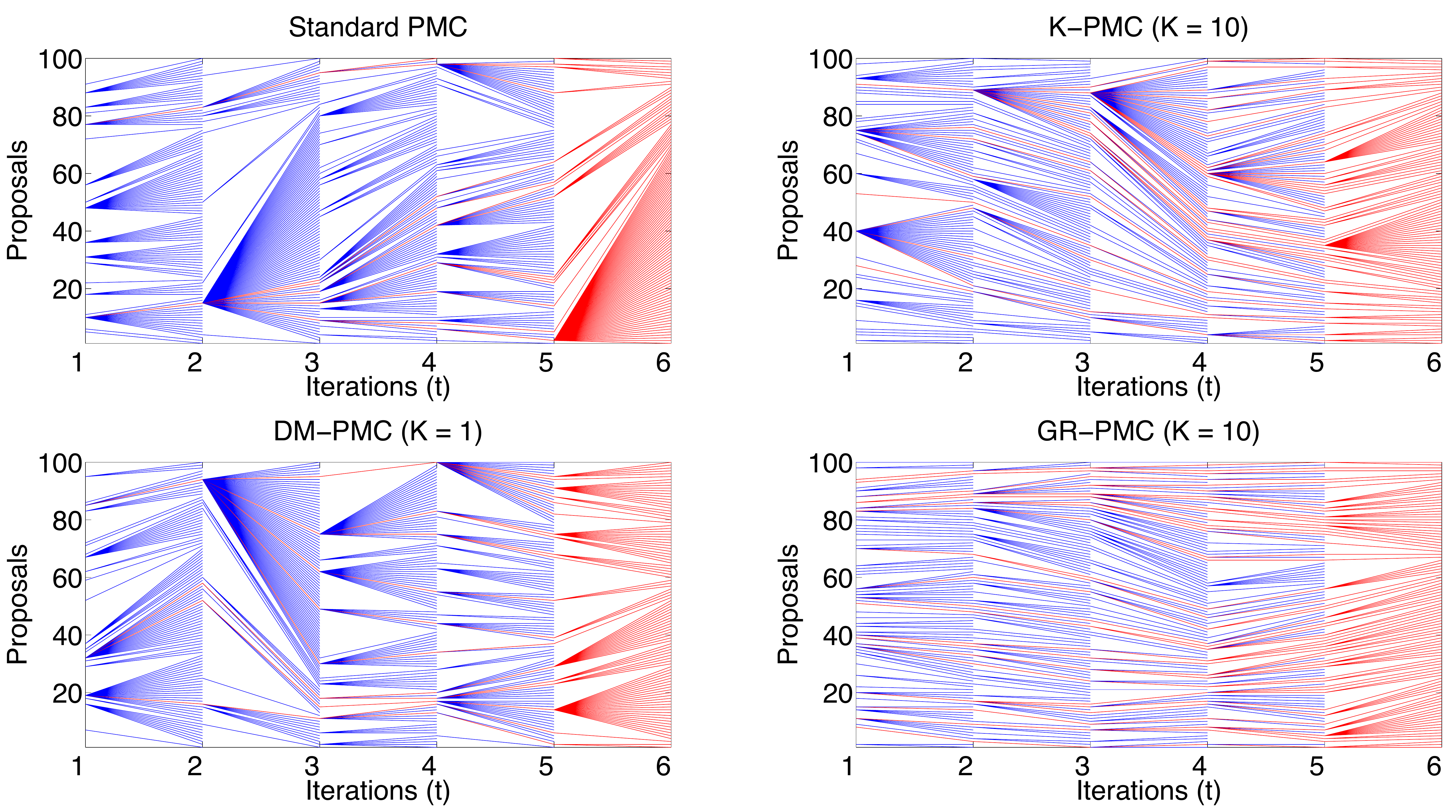}
\caption{{\bf (Ex. of Section \ref{sec_toy_example})} Graphical representation of the indexes of the proposals used to generate the population for the next iteration with different schemes ($6$ iterations; $N=100$, $\sigma=5$). For each pair of iterations, lines link each surviving proposal (``father'' proposal) with the next generation. In red, proposals surviving from the 1st to the 6th iteration.}
\label{figTrellis}
\vspace*{-6pt}
\end{figure*}

Table \ref{table_mean} shows the MSE in the estimation of $E[\X]$ (averaged over both components) for $\x\sim \frac{1}{Z}\pi(\x)$.
We can see that all the proposed schemes outperform the standard PMC for any value of $\sigma$.
In general, the local resampling (LR) works better than the global resampling (GR).
Moreover, we note that the optimum value of $K$ depends on the value of $\sigma$, the scale parameter of the proposals: for small values of $\sigma$ (e.g., $\sigma=1$ or $\sigma=2$) small values of $K$ lead to better performance, whereas a larger value of $K$ (and thus less iterations $T$) can be used for larger values of $\sigma$ (e.g., $\sigma=10$ or $\sigma=20$).
\cblack{In addition, the proposed methods also outperform the M-PMC algorithm in this scenario. Note that M-PMC is an adaptive importance sampling algorithm that does not perform the resampling step.}
{Finally, note that the performance of the SMC sampler can be also improved with the proposed modifications.}

\begin{figure*}
\centering
\subfigure[]{\includegraphics[width=0.49\textwidth]{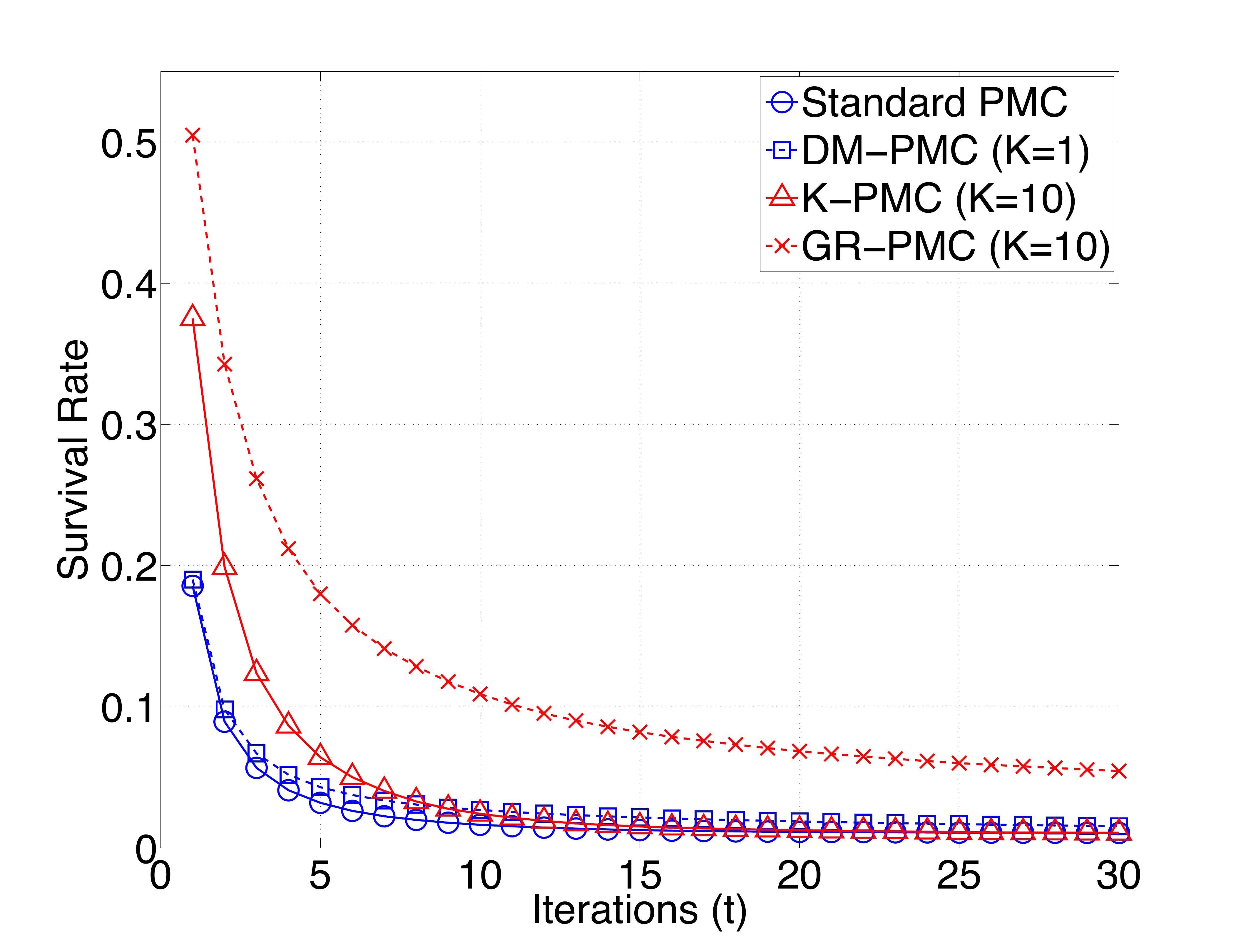}}
\subfigure[]{\includegraphics[width=0.49\textwidth]{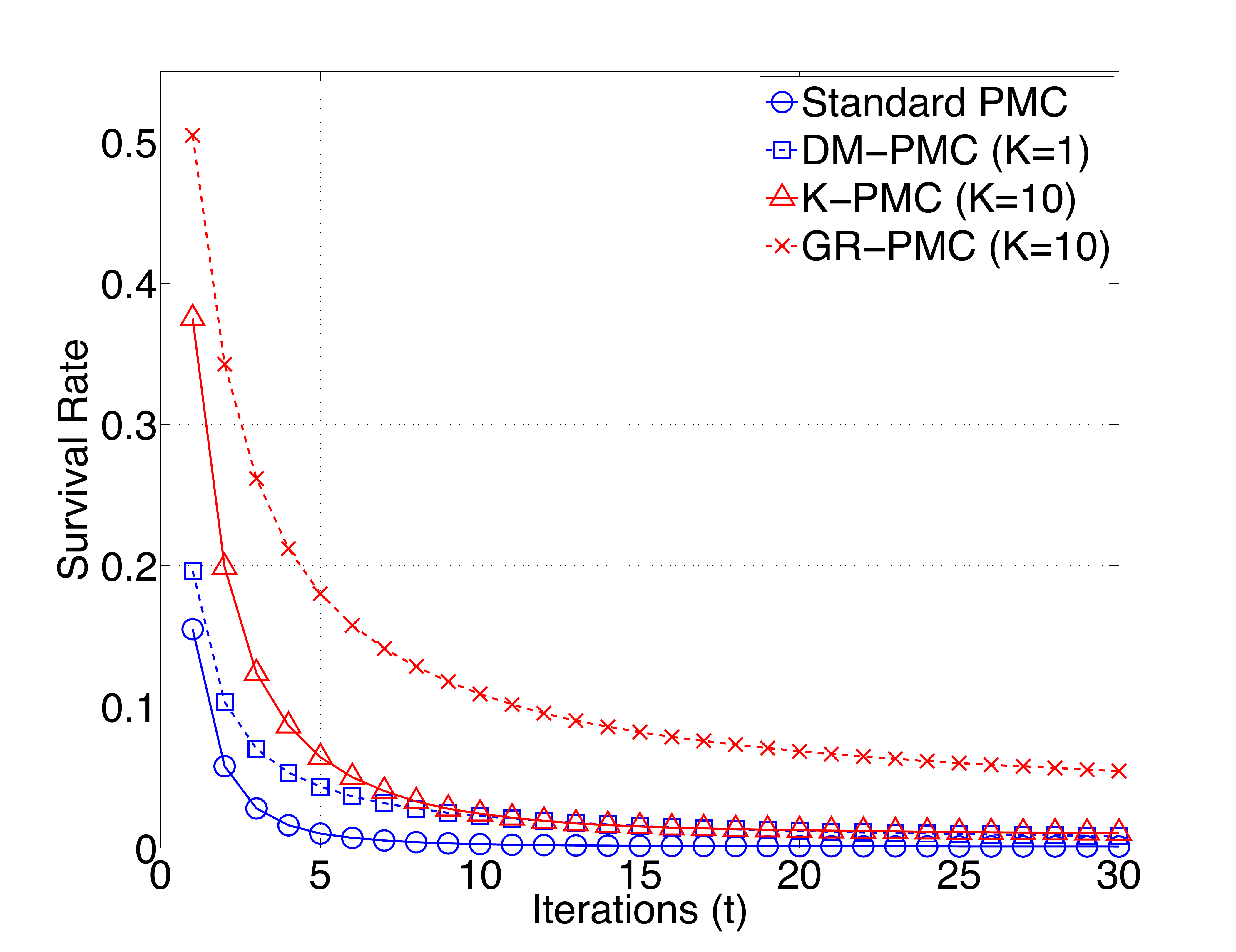}} 
\caption{{\bf (Ex. of Section \ref{sec_toy_example})} Survival rate (after resampling) of the proposals vs. the distance in iterations among the proposals averaged over $500$ runs ($\sigma=5$). {\bf (a)} $N=100$ for all methods. {\bf (b)} Different values of $N$, fixing $NK=1000$ and thus $N \in \{100,1000\}$.}
\label{fig_survival}
\end{figure*}


The large MSE values in Table \ref{table_mean} for some schemes and sets of parameters are due to the fact that they fail at discovering all the modes of the target pdf.
In order to clarify this issue, Fig. \ref{fig_positions_evolutions} shows the evolution of the population of proposals for the first $4$ iterations of the standard PMC ($K=1$), K-PMC (with $K=10$), and DM-PMC with global resampling (also for $K=1$ and $K=10$).
Standard PMC tends to concentrate the whole population on one or two modes, very loosely covering the remaining ones and completely missing the mode in the bottom right corner.
This issue is partly solved by using $K=10$ (after $4$ iterations the proposals are evenly distributed around $3$ out of the $5$ modes) or DM-PMC with $K=1$ (after $4$ iterations the proposals are uniformly distributed among $4$ out of the $5$ modes).
Combining both approaches (DM-PMC and $K=10$) an approximately uniform distribution of the proposals around all the modes of the target is attained.

Finally, in Figs. \ref{figTrellis} and \ref{fig_survival} we explore a well-known problem of PMC: the survival of proposals as the algorithm evolves.
On the one hand, Fig. \ref{figTrellis} shows which proposals have been used to generate the starting population for the next iteration.
After $6$ iterations, all of the $N=100$ proposals in the population have arisen from only $2$ of the proposals in the initial population.
This situation hardly improves by using the DM-PMC: now $4$ initial proposals have generated all the $N=100$ proposals in the $6$-th iteration.
However, by drawing multiple samples per mixand ($K=10$) the situation improves dramatically both when using the standard IS weights ($9$ proposals survive until the $6$-th iteration) and especially when using the DM-PMC ($19$ surviving proposals).
On the other hand, Fig. \ref{fig_survival} shows the evolution in the survival rate of proposals w.r.t. the distance in iterations (or generations). In standard PMC, after very few iterations, most of the ancestors do not survive.
This rate falls down as $t$ increases in all cases, but the DM weights and especially the use of multiple samples per mixand help in slowing down this decrease.
Therefore, we can conclude that the newly proposed schemes can be very useful in preserving the diversity in the population of proposals.

\subsection{High-dimensional example}
\label{sec_high_dimension}

We consider a target corresponding to a mixture of isotropic Gaussians
\begin{equation}
	\pi(\x) = \frac{1}{3}\sum_{k=1}^{3} \mathcal{N}(\x; {\bm \nu}_k, {\bm \Sigma}_k),
\label{eq:hdMixture}
\end{equation}  
where $\x\in \mathbb{R}^{10}$, ${\bm \nu}_k=[\nu_{k,1}, \ldots,\nu_{k,10}]^{\top}$, and ${\bm \Sigma}_k=\xi_k^2 {\bf I}_{10}$ for $k \in \{1,2,3\}$, with $\bI_{10}$ being the $10\times 10$ identity matrix.
We set $\nu_{1,j}=-5$, $\nu_{2,j}=6$, and $\nu_{3,j}=3$ for all $j \in \{1,\ldots,10\}$.
Moreover, we set $\xi_k=8$ for all $k \in \{1,2,3\}$.
The expected value of the target ${\pi}(\x)$ is $E[{X_j}]=\frac{4}{3}$ for $j=1,\ldots,10$, and the normalizing constant is $Z=1$.
 
We use Gaussian proposal densities for all the compared methods.
The initial means (adaptive parameters of the proposals) are selected randomly and independently in all techniques as $\mu_i^{(1)} \sim \mathcal{U}([-6\times 6]^{10})$ for $i=1,\ldots, N$.
We use the same isotropic covariance matrices for all the methods and proposal pdfs, ${\bf C}_i = \sigma^2 \bI_{10}$, and we consider $\sigma \in \{1,5,20\}$.
For every experiment, we run $200$ independent simulations and compute the MSE in the estimation of $E[\X]$ (averaging the MSE of each component).
{We consider the same techniques as in the bi-dimensional example, testing $N\in\{100,1000\}$ and different values of samples per iteration, $K\in \{2,10,20,100\}$.
We have tested different sets of parameters, always keeping the total number of samples fixed to $L = KNT = 2 \cdot 10^5$. Moreover, in this example we implement another variant of the SMC scheme \cite{del2006sequential}, using a sequence of four tempered target densities, $\pi^{(1)}(\x)$, $\pi^{(2)}(\x)$, $\pi^{(3)}(\x)$ and $\pi^{(4)}(\x)=\pi(\x)$. These auxiliary targets have the same form as in Eq. \eqref{eq:hdMixture}, where the diagonal elements of each covariance matrix ${\bm \Sigma}_k^{(s)}$, $s=1,2,3,4$ and $k=1,2,3$, are respectively $\xi_k^{(1)}=16$, $\xi_k^{(2)}=12$, $\xi_k^{(3)}=9$ and, finally, $\xi_k^{(4)}=8$ (the true target). In addition, we also test this algorithm with the residual sampling (see for instance \cite{Cappe05,Douc05}), instead of the standard multinomial resampling. 
}

Table \ref{table_d10} shows that the proposed PMC schemes outperform the standard PMC in most of the cases.
Indeed, a decrease of more than one order of magnitude in the MSE can often be attained by using DM-PMC with an appropriate value of $K$ instead of the standard PMC.
\cblack{Finally, note that, although M-PMC behaves well for most of the parameters tested, overall the proposed methods yield the best performance in terms of MSE and robustness w.r.t. parameter choice.}

\begin{figure}
\centering
\includegraphics[width=0.6\textwidth]{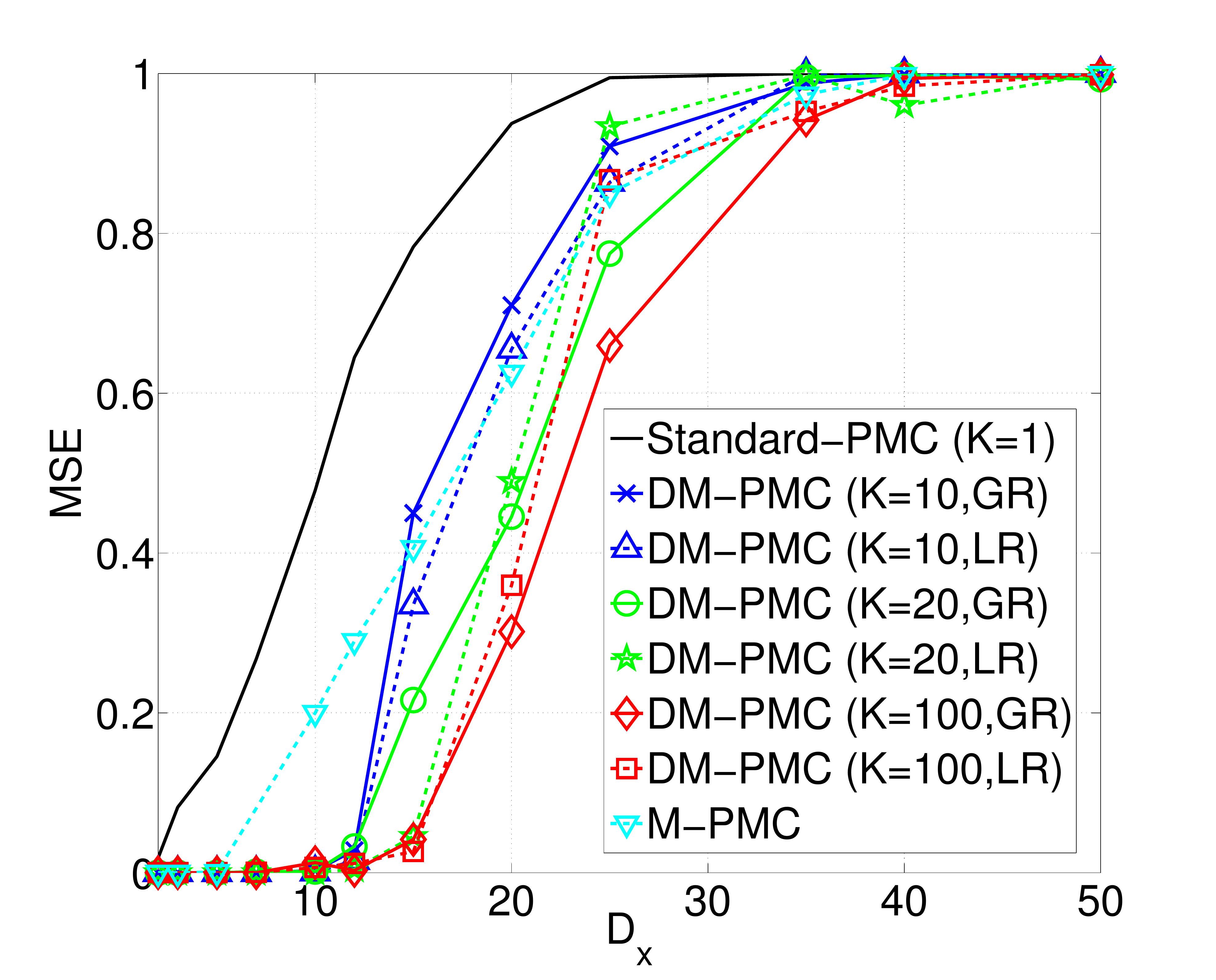}
\caption{{\bf (Ex. of Section \ref{sec_high_dimension})} MSE of the normalizing constant $Z$, using $N=100$ proposals and a scale parameter $\sigma=5$, as the dimension of the state space $D_x$ increases.}
\label{fig_MSE_Z_vs_Dimension}
\end{figure}

In order to study the performance of the proposed schemes as the dimension of the state space increases, we change the dimension of the state space in \eqref{eq:hdMixture}.
Namely, the target density is still a mixture of three isotropic Gaussians with the same structure for the mean vectors and covariance matrices as before, but now the dimension of $\x$ is $D_x \in [1,50]$.
We have tested all the methods with $\sigma = 5$ and $N=100$.
Fig. \ref{fig_MSE_Z_vs_Dimension} shows the evolution of the MSE in the estimation of the normalizing constant as a function of $D_x$.
As expected, the performance of all the methods degrades as the dimension of the problem, $D_x$, becomes larger.
Nonetheless, the performance of the proposed methods decays much more slowly than that of the standard PMC, thus allowing them to still provide a reasonably low MSE in higher dimensions.
Note that, since the true normalizing constant of the target is $Z=1$, when the methods behave poorly in high dimensions and the proposals do not discover the modes, the estimation is $\hat{Z} \approx 0$, and therefore the MSE tends to $1$, which is the worst-case situation.

\begin{landscape}
\begin{table*} 
\setlength{\tabcolsep}{2pt}
\def\marginwidth{1.5mm}
\begin{center}
{\tiny
\begin{tabular}{|l@{\hspace{\marginwidth}}|c@{\hspace{\marginwidth}}|c@{\hspace{\marginwidth}}|c@{\hspace{\marginwidth}}||c@{\hspace{\marginwidth}}|c@{\hspace{\marginwidth}}|c@{\hspace{\marginwidth}}|}
\hline
{\bf Algorithm} & \multicolumn{3}{c| |}{$\bm N=100$} & \multicolumn{3}{c|}{$\bm N=1000$} \\
\cline{2-7}
&  ${\bm \sigma{\bf=1}}$ &  ${\bm \sigma{\bf=5}}$ & ${\bm \sigma{\bf=20}}$ &  ${\bm \sigma{\bf=1}}$ &  ${\bm \sigma{\bf=5}}$ & ${\bm \sigma{\bf=20}}$  \\
\hline
\hline
{Standard PMC \cite{Cappe04}} &12.43 (10.85-14.19) &8.11 (6.47-9.71) &1.24 (0.94-1.61)  &12.68 (9.78-16.14) &5.94 (3.14-10.48) &0.53 (0.32-0.85)\\
\hline
\hline
GR-PMC ($K=2$)&14.53 (13.29-16.07) &4.05 (2.52-6.24) &0.50 (0.43-0.58) &11.90 (7.86-17.65) &\textbf{0.01} (0.01-0.02) &0.15 (0.12-0.20)\\
LR-PMC ($K=2$) &11.55 (9.11-14.29) &12.77 (9.21-15.36) &78.31 (67.67-86.79) &\textbf{2.52} (1.69-3.39) &0.82 (0.50-1.27) &29.44 (20.52-37.92)\\
\hline
GR-PMC ($K=10$) &13.02 (11.69-14.48) &0.91 (0.48-1.58) &0.22 (0.20-0.24)&3.57 (1.82-6.37) &0.10 (0.00-0.27) &0.19 (0.14-0.25)\\
LR-PMC ($K=10$)&8.15 (6.44-10.81) &0.21 (0.13-0.30) &1.85 (1.56-2.12) &4.34 (2.62-6.86) &\textbf{0.01} (0.00-0.01) &1.61 (1.06-2.12)\\
\hline
GR-PMC ($K=20$) &10.89 (9.82-11.92) &0.74 (0.35-1.32) &0.23 (0.20-0.26) &5.45 (2.49-9.43) &0.05 (0.02-0.09) &0.12 (0.08-0.16)\\
LR-PMC ($K=20$) &6.92 (5.56-8.35) &\textbf{0.16} (0.11-0.25) &0.77 (0.68-0.87) &4.59 (2.15-8.21) &0.04 (0.02-0.08) &0.55 (0.42-0.65)\\
\hline
GR-PMC ($K=100$) &7.61 (6.57-8.60) &\textbf{0.16} (0.10-0.29) &\textbf{0.17} (0.15-0.18) &5.71 (3.28-9.78) &0.65 (0.15-1.46) &0.10 (0.07-0.14)\\
LR-PMC ($K=100$) &7.05 (4.99-8.73) &0.41 (0.09-0.99) &0.28 (0.24-0.33) &5.48 (3.01-9.12) &0.17 (0.10-0.28) &0.19 (0.15-0.23)\\
\hline                                                                    
\hline
{\cblack{M-PMC \cite{Cappe08}}} &10.78 (9.53-19.78) &9.06 (4.40-12.72) &0.35 (0.20-0.64)&3.28 (2.77-4.88) &0.12 (0.07-0.50) &\textbf{0.07} (0.05-0.12)\\
\hline 
\hline
{{SMC \cite{del2006sequential}}}&4.99 (3.40-6.87) &0.92 (0.67-1.12) &0.45 (0.35-0.58) &11.45 (7.39-15.67) &1.75 (1.20-2.50) &0.38 (0.25-0.54)\\
{{SMC with tempering \cite{del2006sequential}}} &3.80 (2.76-4.90) &0.56 (0.48-0.65) &0.41 (0.29-0.50) &7.04 (4.75-9.82) &1.64 (1.12-2.03) &0.51 (0.41-0.67)\\ 
{{SMC with tempering and residual resampling \cite{del2006sequential}}} &\textbf{2.79} (2.53-3.15) &0.54 (0.50-0.57) &0.26 (0.24-0.27)  &7.29 (6.73-7.83) &1.24 (1.16-1.35) &0.43 (0.40-0.47)\\ 
\hline 
\end{tabular}
}
\end{center}
\caption{{\bf (Ex. of Section \ref{sec_high_dimension})} MSE in the estimation of $E[\X]$, for $\sigma \in \{1,5,20\}$ and $K \in \{2,10,20,100\}$, keeping the total number of evaluations of the target fixed to $L = 2 \cdot 10^5$.  The dimension space of the target is $D_x=10$. The best results for each value of $\sigma$ are highlighted in bold-face.}
\label{table_d10}
\end{table*}
\end{landscape}

\subsection{Autoregressive filter with non-Gaussian noise}
\label{ARExSect}

We consider the use of an autoregressive (AR) model contaminated by a non-Gaussian noise.
This kind of filters is often used for modeling some financial time series {(see for instance \cite[Section 5]{shumway2013time} and \cite{hamilton1994autoregressive})}, where the noise is assumed to follow the so-called {\it generalized hyperbolic distribution} 
\cite{eberlein2001application}.
Namely, we consider  the following observation model, 
\begin{eqnarray}
\label{ARmodel}
y_m = x_1y_{m-1} + x_2y_{m-2} +x_3y_{m-3} +x_4y_{m-4}+u_m,
\end{eqnarray}
where $m=1,\ldots,M$ is a time index, and $u_m$ is a heavy-tailed driving noise:
\begin{equation*}
	u_m \sim p(u) \propto  e^{\beta(u-\mu)} \frac{B_{\lambda-\frac{1}{2}}\left(\alpha\sqrt{\delta^2+(u-\mu)^2}\right)}{\left(\sqrt{\delta^2+(u-\mu)^2}\right)^{\frac{1}{2}-\lambda}},
\end{equation*}
where $B_{\lambda}$ denotes the \emph{modified Bessel function} \cite{abramowitz1972handbook}.
The vector of unknowns, $\x^* = [x_1^*, x_2^*, x_3^*, x_4^*]^{\top}$, contains the coefficients of the AR model. 

Given a set of observations ${\bf y}=[y_1,\ldots,y_M]^{\top}$, the inference problem consists of obtaining statistical information about  $\x^*$, by studying the corresponding posterior distribution $\normalized{\pi}(\x|{\bf y})$.
More specifically, we have synthetically generated $M=200$ observations, ${\bf y}=[y_1,\ldots,y_M]^{\top}$, setting $\x^*  =  [0.5, ~0.1, ~ -0.8, ~0.1]^\top$, $\lambda=0.5$, $\alpha=2$, $\beta=1$, $\mu=-1$, and $\delta=1$.\footnote{For the generation of i.i.d. samples of the generalized hyperbolic noise, we applied a fast and efficient MCMC technique (the FUSS algorithm \cite{martino2015fast}), drawing samples from univariate distributions. After a few iterations, the resulting samples were virtually independent.}
Assuming improper uniform priors over the unknown coefficients, the objective is computing the expected value ${\bf \hat x}=\int_{\mathbb{R}^4} \x \normalized{\pi}(\x|{\bf y}) d\x$.
{Since we are using $M=200$ observations (a large number for this example), we assume that the posterior pdf is quite sharp and concentrated around the true value, $\x^*  =  [0.5, ~0.1, ~ -0.8, ~0.1]^\top$. Nevertheless, in practice we assume that the inference algorithms have no clue of which is that true value (i.e., we assume no a priori information). Therefore, $\x^*$ is only used for evaluating the performance of the different methods in terms of MSE.}

All the methods use Gaussian proposals, with the initial adaptive parameters of the individual proposals selected uniformly within the $[-6,6]^4$ square, i.e., 
${\bm \mu}_i^{(1)}\sim \mathcal{U}([-6,6]\times[-6,6]\times[-6,6]\times[-6,6])$, and the covariance matrices for all the Gaussians selected as $\bC_i = \sigma^2\bI_4$, with $\sigma =5$ for $i=1,\ldots,N$.
As in the previous examples, we have tested different combinations of parameters, keeping the total number of evaluations of the target fixed to $L=NKT =  2 \cdot 10^5$.
We have evaluated different values of $N\in \{100, 1000, 5000\}$ and $K\in \{5,10,50,100\}$. 
We ran 500 independent simulations and computed the MSE in the estimation of ${\bf \hat x}$ w.r.t. the true value $\x^*$.

The results obtained by the different methods, in terms of MSE averaged over all the components of $\x$, are shown in Table \ref{TableExAR}.
Note that some combinations of $K$ and $N$ would yield a number of iterations $T<1$, since we set $T = L/(NK) =  2 \cdot 10^5/(NK)$. Therefore, those simulations cannot be performed and are indicated in the Table with the symbol $*$.
Note that, for any choice of $N$, the alternative schemes proposed in the paper largely outperform the standard PMC.
Furthermore, the advantage of using $K>1$ can again be clearly seen for the three values of $N$ tested.
More specifically, the smallest the value of $N$ the largest the value of $K$ that should be used to attain the best results.
\cblack{Note also that M-PMC behaves particularly well in this scenario for high values of $N$, but its performance is very poor for $N=100$ (unlike GR-PMC and LR-PMC, which can still provide a good performance for the right value of $K$).}

\begin{table} 
\setlength{\tabcolsep}{2pt}
\def\marginwidth{1.5mm}
\begin{center}
\scriptsize{
\begin{tabular}{|l@{\hspace{\marginwidth}}||c@{\hspace{\marginwidth}}|c@{\hspace{\marginwidth}}|c@{\hspace{\marginwidth}}|}
\hline
\cline{2-4}
{\bf Algorithm} &  ${\bm N{\bf=100}}$ &  ${\bm N{\bf=1000}}$ & ${\bm N{\bf=5000}}$  \\
\hline
\hline
{\small Standard PMC \cite{Cappe04}}  &13615.95 (13197.39-15021.45) &69.99 (62.23-72.24) &0.56 (0.56-0.68)\\
\hline
\hline
GR-PMC ($K=5$) &1597.57 (1516.26-1727.82) &1.92 (1.72-2.22) &0.08 (0.07-0.10)\\
LR-PMC ($K=5$) &31.04 (28.99-33.33) &0.36 (0.33-0.40) &0.20 (0.15-0.24)\\
\hline
GR-PMC ($K=10$) &520.62 (472.44-558.27) &0.30 (0.26-0.40) &0.07 (0.05-0.10)\\
LR-PMC ($K=10$) &14.99 (14.06-15.99) &0.29 (0.28-0.32) &0.21 (0.14-0.27)\\
\hline
GR-PMC ($K=50$) &16.91 (15.43-20.42) &\textbf{0.05} (0.04-0.08) &*\\
LR-PMC ($K=50$) &1.89(1.61-2.12) &0.15 (0.14-0.21) &*\\
\hline
GR-PMC ($K=100$) &2.23 (1.74-3.39) &0.10 (0.06-0.13) &*\\
LR-PMC ($K=100$) &0.77 (0.62-0.90) &0.17 (0.09-0.18) &*\\
\hline
\hline
{\small \cblack{M-PMC \cite{Cappe08}}}  &182.10 (64.29-316.59) &{0.07} (0.06-0.09) &\textbf{0.05} (0.04-0.07)\\
\hline                                                                     
\end{tabular}
}
\end{center}
\caption{{\bf (Ex. of Section \ref{ARExSect})} MSE of $E[\X]$ for different values of $K$ and $N$, keeping the total number of evaluations of the target fixed to $L=KNT=  2 \cdot 10^5$. The symbol $*$ indicates combinations where the number of iterations $T<1$, and therefore they cannot be performed.}
\label{TableExAR}
\end{table}
\subsection{{Localization problem in a wireless sensor network}}
\label{sec_example_localization}
{

Let us consider a static target in a two-dimensional space. The goal consists on positioning the target within a wireless sensor network using only range measurements acquired by some sensors. This example appears in the signal processing literature for localization applications, e.g. in  \cite{Ali07,Ihler05,MartinoSigPro10,APIS15}
In particular, let $\textbf{X}=[X_1,X_2]^{\top}$ denote the random vector representing the position of the target in $\mathbb{R}^{2}$ plane. 
%
%
The measurements are obtained from $6$ range sensors located at $\textbf{h}_1=[1,-8]^{\top}$, $\textbf{h}_2=[8,10]^{\top}$, $\textbf{h}_3=[-15,-7]^{\top}$, $\textbf{h}_4=[-8,1]^{\top}$, $\textbf{h}_5=[10,0]^{\top}$ and $\textbf{h}_6=[0,10]^{\top}$.
The measurements are related to the target position through the following expression:
\begin{gather}
\label{sistemaejemplo}
\begin{split}
Y_{j,r}=-20\log\left(||{\bf x}-{\bf h}_j ||^2\right)+\Theta_{j}, \quad j=1,\ldots, 6,  \quad  r=1,\ldots,d_y,
\end{split}   
\end{gather}
where $\Theta_{j} \sim \mathcal{N}({\bm \theta_j}|{\bf 0},\omega_j^2{\bf I})$, with $\omega_j=5$ for all $j \in 1,\ldots,6$. Note that the total number of data is $6d_y$. We consider a wide Gaussian prior pdf with mean $[0, 0]^{\top}$ and covariance matrix $[\omega_0^2 \ 0; 0 \  \omega_0^2]^{\top}$  with $\omega_0=10$,

We simulate  $6d_y=360$  measurements from  the model ($d_y=60$ observations from each sensor), fixing $x_1=3.5$ and $x_2=3.5$. 
The goal consists in approximating the mean of the posterior distribution ${\bar \pi}({\bf x}|{\bf y})$, through the improved PMC techniques proposed in this paper. In order to compare the different techniques, we computed the value of interest by using an extremely thin grid, yielding $E[{\bf X}]\approx [3.415, 3.539]^{\top}$. 

We test the proposed methods and we compare them with the standard PMC \cite{Cappe04} and the M-PMC \cite{Cappe08}. In all cases, Gaussian proposals are used, with initial mean parameters selected uniformly within the $[1,5]\times[1,5]$ square, i.e., ${\bm \mu}_i^{(1)}\sim \mathcal{U}([-1,5]\times[-4,4])$ for $i=1,\ldots,N$.
All the methods use the same isotropic covariance matrices for all the Gaussian proposals, $\bC_i = \sigma^2\bI_2$ with $\sigma \in \{1,2\}$. We have tried $N\in\{100,500\}$ proposals. In the proposed methods, we test the values $K \in \{20,50,100,200\}$. Note that again, we keep fixed the total number of evaluations to $L=KNT=2 \cdot 10^5$.

 Table \ref{table_sensors} shows the MSE in estimation of the expected value of the posterior, with the different PMC methods. Again, the proposed methods largely beat the standard PMC for all the sets of parameters. The M-PMC algorithm is again competitive (especially with $N=500$), but the proposed algorithms obtain better performance (in particular, the LR-PMC with a high $K$).

}

\begin{table} 
\setlength{\tabcolsep}{2pt}
\def\marginwidth{1.5mm}
{
\begin{center}
\scriptsize{
\begin{tabular}{|l@{\hspace{\marginwidth}}||c@{\hspace{\marginwidth}}|c@{\hspace{\marginwidth}}|c@{\hspace{\marginwidth}}|c@{\hspace{\marginwidth}}|}
\hline
\cline{2-5}
&   \multicolumn{2}{c| |}{${\bm N{\bf=100}}$} &   \multicolumn{2}{c| |}{${\bm N{\bf=500}}$} \\
\cline{2-5}
{\bf Algorithm} &  $\sigma=1$ &  $\sigma=2$ &$\sigma=1$  & $\sigma=2$\\
\hline
\hline
{\small Standard PMC \cite{Cappe04}} &621.85 (542.98-685.76) &2424.35 (1916.39-2995.05) &167.52 (33.10-376.49) &756.46 (490.36-1077.76)\\
\hline
\hline
GR-PMC ($K=20$)  &7.51 (5.83-8.97) &28.02 (23.15-33.41) &0.87 (0.11-1.86) &9.30 (3.33-14.30)\\
LR-PMC ($K=20$)  &0.59 (0.50-0.68) &1.27 (0.89-1.65) &0.25 (0.10-0.54) &0.40 (0.35-0.45)\\
\hline
GR-PMC ($K=50$)  &1.82 (1.50-2.26) &7.00 (5.30-8.56) &0.52 (0.39-0.69) &1.72 (0.19-3.56)\\
LR-PMC ($K=50$)  &0.37 (0.32-0.44) &0.88 (0.70-1.04) &0.25 (0.13-0.32) &0.38 (0.30-0.47)\\
\hline
GR-PMC ($K=100$)  &0.74 (0.63-0.88) &1.66 (1.31-2.05) &0.32 (0.17-0.48) &0.80 (0.51-0.99)\\
LR-PMC ($K=100$)  &0.28 (0.25-0.33) &0.48 (0.39-0.58) &0.23 (0.14-0.33) &0.11 (0.11-0.11)\\
\hline
GR-PMC ($K=200$)  &0.43 (0.36-0.51) &0.57 (0.45-0.66) &0.36 (0.22-0.48) &0.23 (0.01-0.35)\\
LR-PMC ($K=200$)  &0.26 (0.23-0.29) &0.35 (0.28-0.42) &0.16 (0.12-0.20) &0.37 (0.37-0.37)\\
\hline
\hline
{\small \cblack{M-PMC \cite{Cappe08}}}  &7.75 (6.76-7.73) &32.77 (28.34-37.19) &1.07 (0.82-1.33) &1.66 (1.48-1.84) \\
\hline                                                                     
\end{tabular}
}
\end{center}
}
\caption{{{\bf (Ex. of Section \ref{sec_example_localization}) MSE of the estimator of $E[\X]$ with different PMC algorithms.}}}
\label{table_sensors}
\end{table}

\section{Conclusions}
\label{sec_conclusions}

The population Monte Carlo (PMC) method is a well-known and widely used scheme for performing statistical inference in many signal processing problems.
Three improved PMC algorithms are proposed in this paper.
All of them are based on the deterministic mixture (DM) approach, which provides estimators with a reduced variance (as proved in this paper) and increases the exploratory behavior of the resulting algorithms.
Additionally, two of the methods draw multiple samples per mixand (both with local and global resampling strategies) to prevent the loss of diversity in the population of proposals.
The proposed approaches are shown to substantially outperform the standard PMC on three numerical examples. The proposed improvements can be applied to other existing PMC implementations and other importance sampling techniques, to achieve similar benefits.

\section*{Acknowledgment}

This work has been supported by the Spanish government's projects AGES (S2010/BMD-2422), ALCIT (TEC2012-38800-C03-01), COMPREHENSION \newline(TEC2012-38883-C02-01), DISSECT (TEC2012-38058-C03-01), and OTOSiS (TEC2013-41718-R); by the BBVA Foundation through project MG-FIAR (``I Convocatoria de Ayudas Fundaci\'on BBVA a Investigadores, Innovadores y Creadores Culturales''); by ERC grant 239784 and AoF grant 251170; by the National Science Foundation under Award CCF-0953316; and by the European Union's 7th FP through the Marie Curie ITN MLPM2012 (Grant No. 316861).

\section*{References}

\appendix

\section{Standard vs. deterministic mixture importance sampling}
\label{sec_DM_appendix}
{\color{black}
In this appendix we review the IS estimators, analyzing the properties (unbiasedness and variance) of the estimator with the DM weights. For the sake of clarity, we remove the temporal indexes.
\subsection{Importance sampling estimators}
\label{weights}
Let us consider the estimator of Eq. \eqref{eq_partial_estimator_unnorm} when we have a set of $N$ proposal pdfs, $\{q_i(\x)\}_{i=1}^N$. We draw exactly $K_i=1$ sample from each proposal, i.e., $\x_i \sim q_i(\x)$ for $i=1,\ldots,N$. \bfootnote{From now on, we use $K_i=1$, with $i=1,...,N$, for the sake of clarity, but the analysis can be straightforwardly extended to any $K_i$.}
If the normalizing constant $Z$ is known, the IS estimator is then
\begin{equation}
	\hat{I} = \frac{1}{NZ}\sum_{i=1}^{N} {w_{i} f(\x_{i})}.
\label{eq_estimator_unnormalized}
\end{equation}
The difference between the standard and deterministic mixture (DM) IS estimators lies in the computation of the unnormalized weights.
On the one hand, we recall the standard IS weights are given by
\begin{equation}
	w_{i} = \frac{\pi(\x_{i})}{q_i(\x_{i})},
\label{eq:weightsIS}
\end{equation}
where $\pi(\x_{i})$ is the target evaluated at the $i$-th sample (drawn from the $i$-th proposal).
Substituting \eqref{eq:weightsIS} into  \eqref{eq_estimator_unnormalized}, we obtain the standard IS estimator,
\begin{equation}
	\hat{I}_{IS} = \frac{1}{N  Z}\sum_{i=1}^{N}{{\frac{f(\x_{i}) \pi(\x_{i})}{q_i(\x_{i})}}}.
\label{eq_IS_Est_unnormalized}
\end{equation}
On the other hand, the weights in the DM approach are given by
\begin{equation}
	w_{i} = \frac{\pi(\x_{i})}{\sum_{j=1}^{N}{q_j(\x_{i})}}.
\label{eq:weightsDM}
\end{equation}
Substituting \eqref{eq:weightsDM} into \eqref{eq_estimator_unnormalized} we obtain the DM estimator
\begin{equation}
	\hat{I}_{DM} = \frac{1}{N  Z}\sum_{i=1}^{N}{{\frac{f(\x_{i}) \pi(\x_{i})}{\frac{1}{N} \sum_{j=1}^N q_j(\x_{i})}}}.
\label{eq_DM_Est_unnormalized}
\end{equation}
}

\subsection{\cblack{Unbiasedness of the DM-IS estimator}}
\label{DM_unbiased_appendix}
{\color{black}
It is well known that $\hat I_{IS}$ in Eq. \eqref{eq_IS_Est_unnormalized} is an unbiased estimator of the integral $I$ define in Eq. \eqref{eq:integral} \cite{Robert04,Liu04b}. In this section, we prove that the DM-IS estimator in Eq. \eqref{eq_DM_Est_unnormalized} is also unbiased. Since $\x_i \sim q_i(\x)$, we have
\cblack{
\begin{eqnarray}
E[\hat{I}_{DM}]  &=& \frac{1}{NZ} \sum_{i=1}^N  E_{q_i}\left[\frac{f(\X_i) \pi(\X_i)}{\frac{1}{N}\sum_{j=1}^{N}{q_j(\X_i)}}\right]\\
&=& \frac{1}{NZ} \sum_{i=1}^N  \int \frac{f(\x_i) \pi(\x_i)}{\frac{1}{N}\sum_{j=1}^{N}{q_j(\x_i)}} q_i(\x_i)d\x_i \label{eq_i_2}\\
&=&   \frac{1}{Z} \int \frac{f(\x) \pi(\x)}{\frac{1}{N}\sum_{j=1}^{N}{q_j(\x)}}  \left[\frac{1}{N}\sum_{i=1}^N q_i(\x)\right]d\x \label{eq_i_3}\\
&=&  \frac{1}{Z} \int f(\x) \pi(\x) d\x = I.
\end{eqnarray}
$\hfill\qed$
}

}
{\color{black}
\subsection{\cblack{Variance of the DM-IS estimator}}
\label{DM_variance_appendix}
In this section, we prove that the DM-IS estimator in Eq. \eqref{eq_DM_Est_unnormalized} always has  a lower or equal variance than the standard IS estimator of  Eq. \eqref{eq_IS_Est_unnormalized}. We also consider the standard mixture (SM) estimator $\hat{I}_{SM}$, where $N$ samples are independently drawn from the mixture of proposals, i.e., ${\bf z}_i \sim \frac{1}{N}\sum_{j=1}^{N}q_j(\x)$, and
\begin{equation}
	\hat{I}_{SM}  = \frac{1}{NZ} \sum_{i=1}^N  \frac{f({\bf z}_i) \pi({\bf z}_i)}{\frac{1}{N}\sum_{j=1}^{N}{q_j({\bf z}_i)}}. 
\label{eq:estSM}
\end{equation}
%
%
Note that obtaining an IS estimator with finite variance essentially amounts to having a proposal with heavier tails than the target. See \cite{Robert04,Geweke89} for sufficient conditions that guarantee this finite variance.
\begin{theorem}
For any target distribution, $\pi(\x)$, any square integrable function {w.r.t. $\pi(\x)$}, $f(\x)$, and any set of proposal densities, $\{ q_i(\x) \}_{i=1}^N$, such that the variance of the corresponding estimators is finite,
the variance of the DM estimator is always lower or equal than the variance of the corresponding standard IS and mixture (SM) estimators, i.e., 
\begin{equation}
	\textrm{Var}(\hat{I}_{DM}) \le \textrm{Var}(\hat{I}_{SM}) \le \textrm{Var}(\hat{I}_{IS}).
\label{eq:varIneq}
\end{equation}
\label{theorem_1}
\end{theorem}

\noindent \textit{Proof:}  
The proof is given by Proposition \ref{prop_1} and Proposition \ref{prop_2}.\qed 

{\color{black}
\begin{proposition}
\begin{equation}
	\textrm{Var}(\hat{I}_{SM}) \le \textrm{Var}(\hat{I}_{IS}).
\end{equation}
\label{prop_1}
\end{proposition}
\noindent \textit{Proof:}  
The variance of the IS estimator is given by
\begin{equation}
	\textrm{Var}(\hat{I}_{IS})  = \sum_{i=1}^N  \frac{1}{N^2Z^2}\int \frac{f^2(\x) {\pi}^2(\x)}{q_i(\x)}d\x - \frac{I^2}{N},
\label{eq:varIS}
\end{equation}
where $I =  \frac{1}{Z}\int f(\x) {\pi}(\x) d\x$ is the true value of the integral that we want to estimate \cite{elvira2015generalized}. The variance of the SM estimator is given by
\begin{eqnarray}
	\textrm{Var}(\hat{I}_{SM})  &=& \frac{1}{N^2} \sum_{i=1}^N  \left(\frac{1}{Z^2}\int  \frac{f^2(\x) {\pi}^2(\x)}{\psi(\x)}d\x - I^2\right) \nonumber\\
	&=& \frac{1}{NZ^2}\int  \frac{f^2(\x) {\pi}^2(\x)}{\psi(\x)}d\x - \frac{I^2}{N},
\label{eq:varSM}
\end{eqnarray}
where $\psi(\x)=\frac{1}{N}\sum_{j=1}^{N}{q_j(\x)}$.
Substracting \eqref{eq:varSM} and \eqref{eq:varIS}, we get
\begin{align*}
&	\textrm{Var}(\hat{I}_{SM}) - \textrm{Var}(\hat{I}_{IS}) = \frac{1}{N^2Z^2} \int{\left(\frac{N}{\frac{1}{N}\sum_{j=1}^{N}{q_j(\x)}} - \sum_{i=1}^{N}{\frac{1}{q_i(\x)}} \right)
	 	f^2(\x) {\pi}^2(\x) d\x}.
\end{align*}
Hence, since $f^2(\x) {\pi}^2(\x) \ge 0 \ \forall \x$, in order to prove the theorem it is sufficient to show that
\begin{equation}
	\frac{1}{\frac{1}{N}\sum_{j=1}^{N}{q_j(\x)}} \le \frac{1}{N} \sum_{i=1}^{N}{\frac{1}{q_i(\x)}}.
\label{eq:varIneq2}
\end{equation}
Now, let us note that the left hand side of \eqref{eq:varIneq2} is the inverse of the arithmetic mean of $q_1(\x),\ \ldots,\ q_N(\x)$,
\begin{equation*}
	A_N = \frac{1}{N}\sum_{j=1}^{N}{q_j(\x)},
\end{equation*}
whereas the right hand side of \eqref{eq:varIneq2} is the inverse of the harmonic mean of $q_1(\x),\ \ldots,\ q_N(\x)$,
\begin{equation*}
	\frac{1}{H_N} = \frac{1}{N} \sum_{i=1}^{N}{\frac{1}{q_i(\x)}}.
\end{equation*}
Therefore, the inequality in \eqref{eq:varIneq2} is equivalent to stating that $\frac{1}{A_N} \le \frac{1}{H_N}$, or equivalently $A_N \ge H_N$, which is the well-known arithmetic mean--harmonic mean inequality for positive real numbers \cite{hardy1952inequalities,abramowitz1972handbook}.
{Note that \eqref{eq:varIneq2} can also be proved using Jensen's inequality in Eq. \eqref{eq:jensen} with $\varphi(x) = \frac{1}{x}$, $\alpha_i=\frac{1}{N}$ and $z_i = q_i(\x)$ for $i=1,\ldots,N$.} $\hfill \Box$
}
{\color{black}
\begin{proposition}
\begin{equation}
	\textrm{Var}(\hat{I}_{DM}) \le \textrm{Var}(\hat{I}_{SM}).
\end{equation}
\label{prop_2}
\end{proposition}
\noindent \textit{Proof:}  
The variance of $\hat{I}_{DM}$ is computed 
\begin{eqnarray}
	Var(\hat{I}_{DM})  &=& \frac{1}{N^2Z^2} \sum_{i=1}^N  \left(E_{q_i}\left[ \frac{f^2(\X_i) {\pi}^2(\X_i)}{\psi^2(\X_i)} \right]   - E_{q_i}^2\left[ \frac{f(\X_i) {\pi}(\X_i)}{\psi(\X_i)}  \right] \right) \nonumber \\
 &=&\frac{1}{N^2Z^2} \sum_{i=1}^N \left( \int \frac{f^2(\x){\pi}^2(\x)}{\psi^2(\x)}q_i(\x)d\x\right)  - \frac{1}{N^2Z^2} \sum_{i=1}^N \left(\int \frac{f(\x){\pi}(\x)}{\psi(\x)}q_i(\x)d\x \right)^2 \nonumber \\
  &=&\frac{1}{NZ^2} \left( \int \frac{f^2(\x){\pi}^2(\x)}{\psi^2(\x)} \left[ \frac{1}{N} \sum_{i=1}^N q_i(\x) \right]d\x\right)  - \frac{1}{N^2Z^2} \sum_{i=1}^N \left(\int \frac{f(\x){\pi}(\x)}{\psi(\x)}q_i(\x)d\x \right)^2 \nonumber \\
  &=&\frac{1}{NZ^2} \int \frac{f^2(\x){\pi}^2(\x)}{\psi(\x)}d\x - \frac{1}{N^2Z^2} \sum_{i=1}^N \left(\int \frac{f(\x){\pi}(\x)}{\psi(\x)}q_i(\x)d\x \right)^2
\label{eq_var_DM}
\end{eqnarray}
Analyzing Eqs. \eqref{eq:varSM} and \eqref{eq_var_DM}, we see that proving $\textrm{Var}(\hat{I}_{DM}) \le \textrm{Var}(\hat{I}_{SM})$ is equivalent to proving that
\begin{eqnarray}
\frac{1}{Z^2}\sum_{i=1}^N \left(\int \frac{f(\x){\pi}(\x)}{\psi(\x)}q_i(\x) d\x \right)^2 &\geq& NI^2 \nonumber \\
\frac{1}{Z^2}\sum_{i=1}^N \left(\int \frac{f(\x){\pi}(\x)}{\psi(\x)}q_i(\x) d\x \right)^2 &\geq& N \left( \frac{1}{Z} \int \frac{f(\x){\pi}(\x)}{\psi(\x)}\psi(\x) d\x \right)^2 \nonumber \\
\sum_{i=1}^N \left(\int \frac{f(\x) {\pi}(\x)}{\psi(\x)}q_i(\x) d\x \right)^2 &\geq& N \left( \int \frac{f(\x){\pi}(\x)}{\psi(\x)}\left( \frac{1}{N} \sum_{i=1}^N q_i(\x) \right) d\x \right)^2 \nonumber \\
\sum_{i=1}^N \left(\int \frac{f(\x){\pi}(\x)}{\psi(\x)}q_i(\x) d\x \right)^2 &\geq&  \frac{1}{N} \left( \sum_{i=1}^N\int \frac{f(\x){\pi}(\x)}{\psi(\x)}q_i(\x)d\x \right)^2
\label{eq:cs_jensen}
\end{eqnarray}
{By defining $a_i=a_i(\x)=\int \frac{f(\x){\pi}(\x)}{\psi(\x)}q_i(\x)d\x$, \eqref{eq:cs_jensen} can be expressed more compactly as}
\begin{equation}
	N \sum_{i=1}^N a_i^2 \geq \left( \sum_{i=1}^N a_i\right)^2.
\label{eq_cs_proof}
\end{equation}
The inequality in Eq. (\ref{eq_cs_proof}) holds, since it corresponds to the definition of the Cauchy-Schwarz inequality \cite{hardy1952inequalities}, 
 \begin{eqnarray}
\left( \sum_{i=1}^N a_i ^2\right) \left( \sum_{i=1}^N b_i ^2\right) \geq\left( \sum_{i=1}^N a_i b_i \right)^2,
 \end{eqnarray}
with $b_i=1$ for $i=1,...,N$.
{Once more, \eqref{eq:cs_jensen} can also by proved by using Jensen's inequality in \eqref{eq:jensen} with $\varphi(x)=x^2$, $\alpha_i=\frac{1}{N}$ and $z_i=a_i(\x)$ for $i=1,\ldots,N$.} $\hfill\Box$
}
}

\begin{figure*}[htb]

\centering

\subfigure[]{\hspace{-15mm} \includegraphics[width=1.2\textwidth]{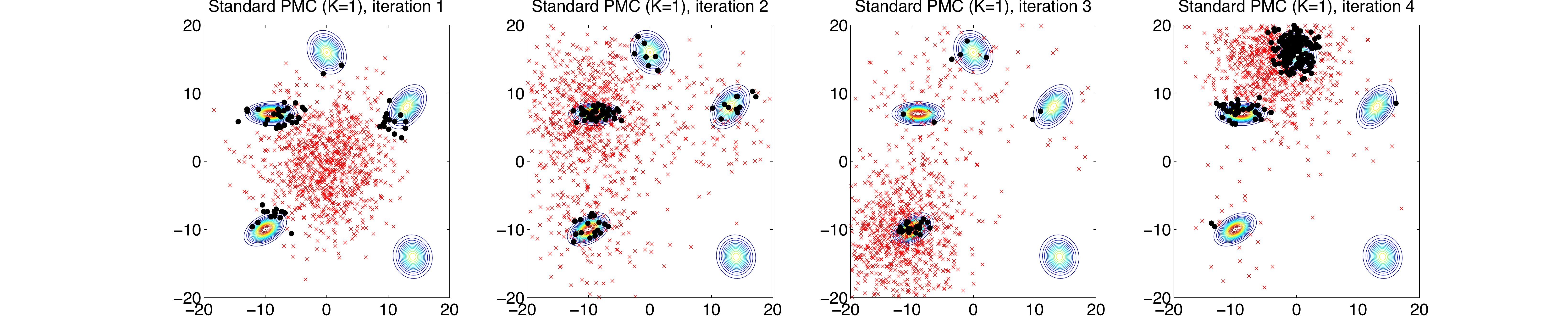}}
\subfigure[]{\hspace{-15mm}\includegraphics[width=1.2\textwidth]{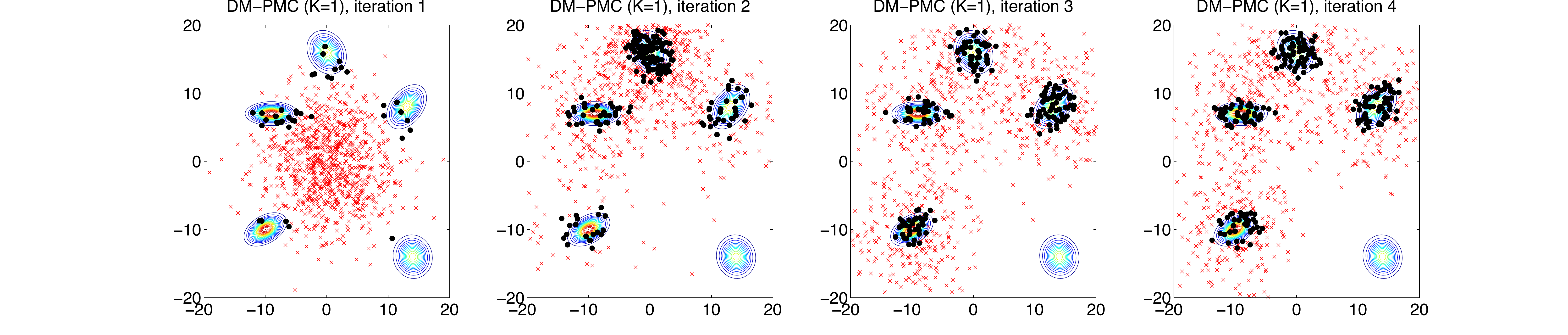}}
\subfigure[]{\hspace{-15mm}\includegraphics[width=1.2\textwidth]{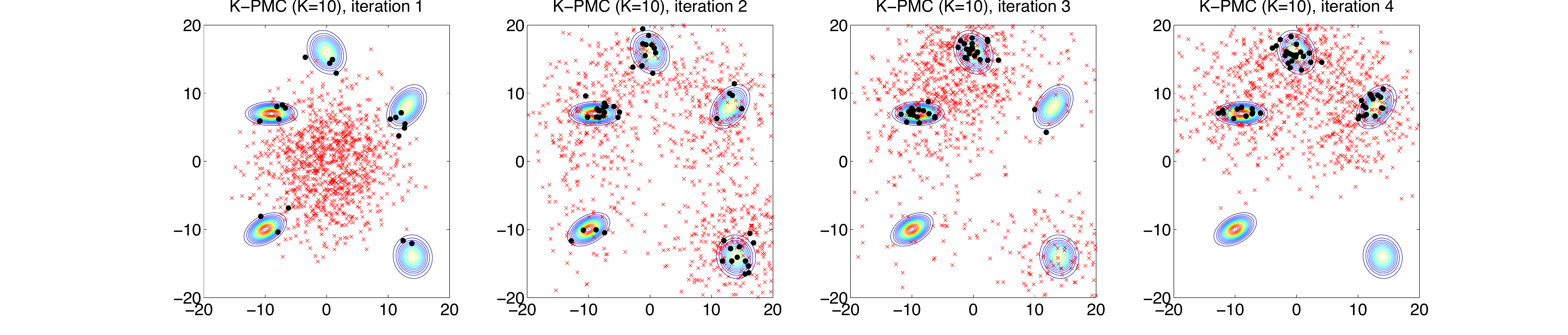}}
\subfigure[]{\hspace{-15mm}\includegraphics[width=1.2\textwidth]{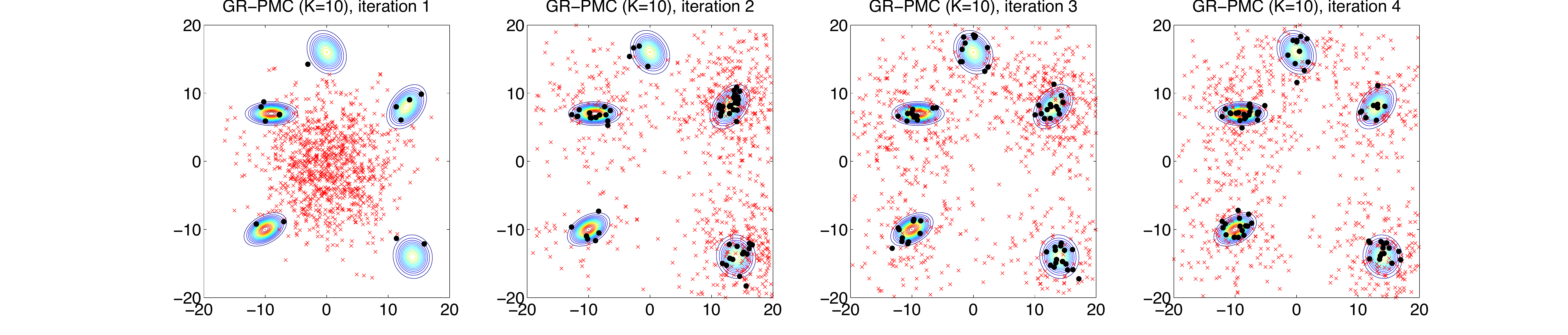}}

\caption{{\bf (Ex1-Section \ref{sec_toy_example})} Evolution of the samples before (red crosses) and after resampling (black circles) for different schemes using $N=100$ and $\sigma=5$. The contour lines of the target density are also depicted. {\bf (a)} Standard PMC. {\bf (b)} DM-PMC (K=1). {\bf (c)} K-PMC (K=10) with global resampling. {\bf (d)} GR-PMC (K=10).}
\label{fig_positions_evolutions}
\end{figure*}
\end{document}